\documentclass{aastex61}
\usepackage{graphicx}
\usepackage{natbib}
\bibpunct{(}{)}{;}{a}{}{,}
\usepackage{amstext}
\usepackage{url}
\usepackage{subfigure}
\usepackage{soul}

\shorttitle{Disorder or order: on radial magnetic fields in young SNRs}
\shortauthors{West et al.}
\received{July 24, 2017}
\revised{October 3, 2017}
\accepted{October 18, 2017}
\submitjournal{ApJL}
\begin{document}

\title{When disorder looks like order: A new model to explain radial magnetic fields in young supernova remnants}

\correspondingauthor{J. L. West}
\email{jennifer.west@dunlap.utoronto.ca}

\author[0000-0001-7722-8458]{J. L. West}
\affil{Dunlap Institute for Astronomy and Astrophysics University of Toronto, Toronto, ON M5S 3H4, Canada}

\author{T. Jaffe}
\affil{CRESST, NASA Goddard Space Flight Center, Greenbelt, MD 20771, USA}
\affil{Department of Astronomy, University of Maryland, College Park, MD , 20742, USA}

\author{G. Ferrand} 
\affil{RIKEN, Astrophysical Big Bang Laboratory, Wako, Saitama-ken, Japan}

\author{S. Safi-Harb}
\affil{Dept of Physics and Astronomy, University of Manitoba, Winnipeg R3T 2N2, Canada}

\author{B. M. Gaensler}
\affil{Dunlap Institute for Astronomy and Astrophysics University of Toronto, Toronto, ON M5S 3H4, Canada}

\begin{abstract} Radial magnetic fields are observed in all known young, shell-type supernova remnants (SNRs) in our Galaxy, including Cas A, Tycho, Kepler, and SN1006 and yet the nature of these radial fields has not been thoroughly explored. Using a 3D model, we consider the existence and observational implications of an intrinsically radial field. We also present a new explanation of the origin of the radial pattern observed from polarization data as resulting from a selection effect due to the distribution of cosmic-ray electrons (CREs). We show that quasi-parallel acceleration can concentrate CREs at regions where the magnetic field is radial, making a completely turbulent field appear ordered, when it is in fact disordered.  We discuss observational properties that may help distinguish between an intrinsically radial magnetic field and the case where it only appears radial due to the CRE distribution. We also show that the case of an intrinsically radial field with a quasi-perpendicular CRE acceleration mechanism has intriguing similarities to the observed polarization properties of SN1006.
\end{abstract}


\section{\label{sec:intro}Introduction}

\begin{figure}
\centering
\includegraphics[width=18cm]{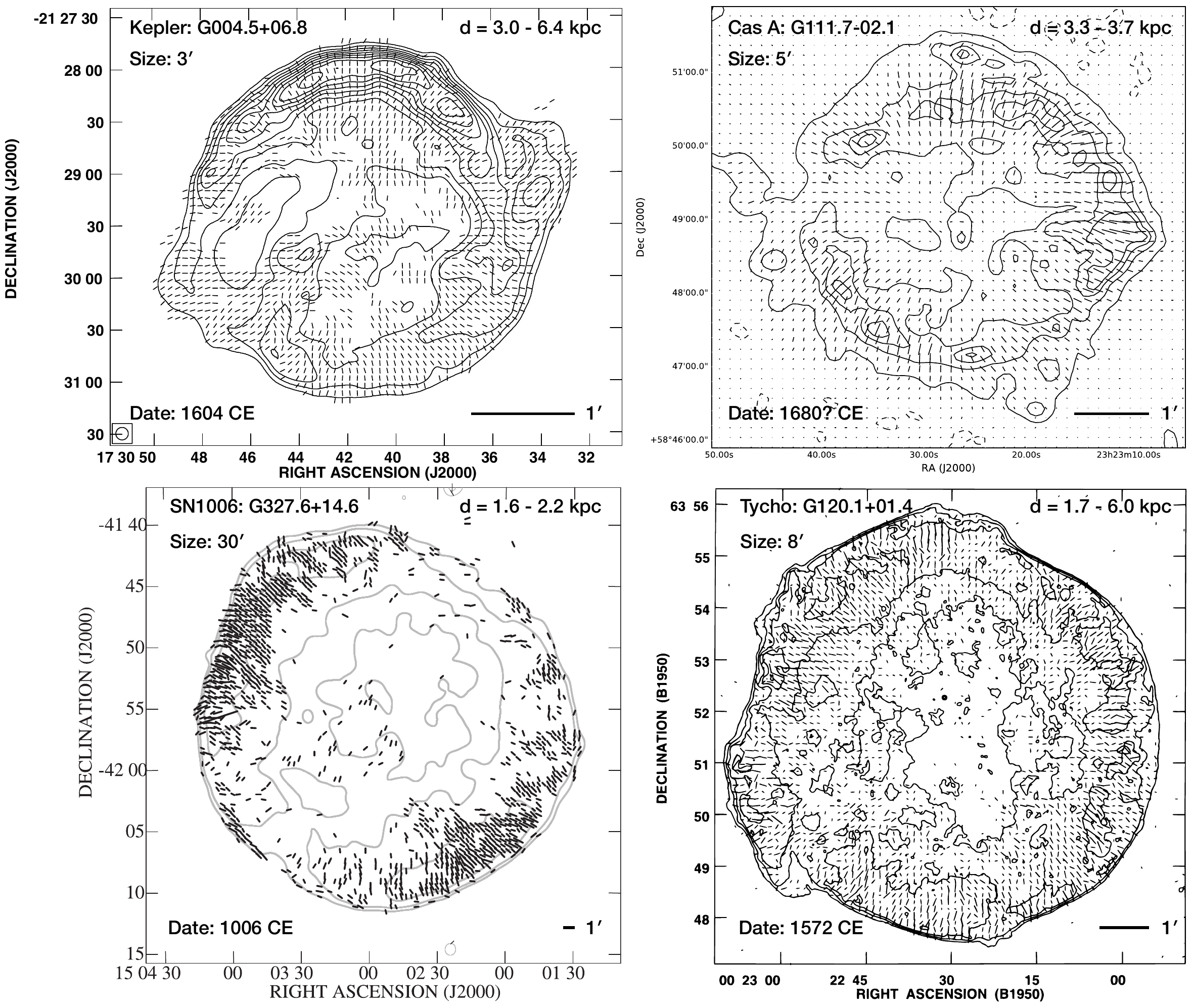}

\caption{\label{fig:historical_snrs}Magnetic fields in young, shell-type SNRs shown with total intensity contours and magnetic field vectors, which have been corrected for Faraday rotation. Top left: G004.5+006.8 (Kepler) at 5.0~GHz \citet[][with permission]{2002ApJ...580..914D}. Top right: G111.7--02.1 (Cas A) at 5.0~GHz (using VLA data provided by L. Rundick). Bottom left: G120.1+01.4 (Tycho) at 1.4~GHz, \citet[][with permission]{1997ApJ...491..816R}. Bottom right: G327.6+14.6 (SN1006) at 1.4~GHz, \citet[][with permission]{Reynoso:2013tr}. Distance and date information is compiled by \citet[][]{Ferrand:2012cr, Green:2014vb} and references therein.}
\end{figure}

There are many outstanding questions surrounding our current understanding of the magnetic fields and cosmic-ray electron (CRE) distribution in supernova remnants (SNRs). Young shell-type SNRs are known to be sites of cosmic-ray acceleration \citep[e.g.,][]{2017arXiv170608275M} and understanding the magnetic field geometry in these SNRs is of particular relevance, since magnetic fields are thought to play a key role in the acceleration mechanism. All historical SNRs observed to date show radial magnetic fields \citep{Milne:1987wi, 1995ApJ...441..300A, 1997ApJ...491..816R, 2002ApJ...580..914D, 2002nsps.conf....1R, Reynoso:2013tr}, the origin of which is still not completely understood \citep[e.g.,][]{2012SSRv..166..231R}. Four such observations are shown in Fig.~\ref{fig:historical_snrs}. 

Due to the ordered radial appearance, a common assumption has been that the magnetic field has some intrinsically radial nature that leads to this appearance. Two possible explanations are: owing to the expansion, Rayleigh-Taylor instabilities will stretch the field lines preferentially along the radial component, and/or turbulence with a radially biased velocity dispersion is induced, leading to selective amplification of the radial component of the magnetic field  \citep{1996ApJ...472..245J,2001ApJ...560..244B,2008ApJ...678..939Z, Inoue:2013el}. However, there are no 3D models that predict the appearance of the various observational properties, such as synchrotron emission and polarization, which such an intrinsically radial field would produce. In addition, it does not seem plausible that the intrinsic radial field would exist without some component of turbulence and the observational consequences of this have not been studied.

SN1006 is the oldest of the SNRs in Fig.~\ref{fig:historical_snrs} and it is the only one to have a distinctly bilateral appearance. The others have a round shape with radio emission appearing all around their shell. Observations of both SN1006 and Cas~A show evidence of a tangential field running along the outer edge of the shell \citep{Reynoso:2013tr, 2001ApJ...552L..39G}. In addition to being the oldest and only bilateral SNR, SN1006 also differs from the others as it is also the largest (both physical and angular size) and it is located at the highest Galactic latitude where the ambient Galactic medium may be more uniform and the Galactic magnetic field may be more ordered \citep{2015ASSL..407..483H}. This may result in the difference in amplitude between the ordered component and a turbulent component, which must be present at some level, and may explain the aforementioned tangential field at the outer perimeter of the shell.

Since radial fields are observed almost exclusively in young SNRs, it must be assumed that there exists some kind of transition from a radial magnetic field to a tangential one, which is observed in many older SNRs \citep{1976AuJPh..29..435D, 2012SSRv..166..231R}. \citet{2016A&A...587A.148W} \citep[see also][]{1998ApJ...493..781G} found that a model made by compressing the ambient Galactic magnetic field can explain the geometry of most of the bilateral-shaped SNRs, which is consistent with this picture. This model can also explain the morphology of SN1006, but fails to reproduce the radial magnetic field \citep[e.g.,][]{2017A&A...597A.121W}. 

In the case of SN1006, the favoured explanation of its radial-looking field uses quasi-parallel CRE acceleration and a ``polar-cap" model of a compressed,  ordered ambient magnetic field. This SNR has been the subject of much debate \citep[e.g.,][review by \citealt{2017arXiv170202054K}]{Rothenflug:2004ch, Petruk:2009bg, 2011A&A...531A.129B,Schneiter:2015fc,2017A&A...597A.121W,2017MNRAS.466.4851V} involving both its radial magnetic field geometry, its CRE distribution, and how this relates to its bilateral morphology, but these studies do not consider the presence of an intrinsically radial magnetic field. 

Two CRE acceleration scenarios are typically considered: (a) the quasi-perpendicular scenario, where CREs are most efficiently accelerated when the shock normal is perpendicular to the post-shock magnetic field, and (b) the quasi-parallel scenario, where CREs are most efficiently accelerated when the shock normal is parallel to the post-shock magnetic field (\citealp{Jokipii:1982jy,1989ApJ...338..963L,Fulbright:1990gu}
and references therein). These two scenarios have been applied to the geometry of what is assumed to be an ordered, compressed magnetic field resulting in two opposite pictures of the bilateral morphology (rotated from each other by 90$^\circ$). Which, if any, applies has been argued in many studies (see references above) and \citet{2017A&A...597A.121W} conclude that neither scenario can adequately describe all the observed features of SN1006.

We consider the observational properties of an intrinsically radial magnetic field. We also explore the consequences of including a turbulent magnetic field component, which has led us to a new explanation of the origin of radial-appearing polarization observations. We describe our modelling in Sec.~\ref{sec:model}. Results and discussion are presented in Sec.~\ref{sec:result}, and conclusions in Sec.~\ref{sec:conclusions}.

\section{\label{sec:model}Modelling}

We use the updated version of Hammurabi\footnote{\url{https://sourceforge.net/projects/hammurabicode/}} \citep[][]{Waelkens:2009bn}, a HEALPix-based modelling code for simulating 2D synchrotron emission maps from 3D models of the input magnetic field and electron distributions. Hammurabi uses the HEALPIX pixelization scheme \citep{Gorski:2005ku} and integrates along lines of sight to calculate simulated total radio synchrotron emission, Stokes I, and the polarization vectors Stokes Q and Stokes U. 

We use a static, non-dynamical, geometric model for this study. The models have a 512x512x512 pixel grid with a physical size of 21~pc on a side giving a physical scale of 0.04~pc/pixel. The physical size of the grid is arbitrary and does not affect the results, but we choose to use a value that is consistent with the size of the oldest and largest SNR in the sample in Fig.~\ref{fig:historical_snrs}, SN1006 (where the box size is defined as $1.2*d$, where $d=17.5$~pc, which is the diameter of SN1006). We construct a 3D box and define the magnetic field at each point in the box based on some physical assumptions.  We model several linear combinations of the following three components:

\begin{enumerate}
\item An ordered regular component defined by a compressed ambient field, $\mathbf{B}_{\textrm{reg}}$ (see Sec.~\ref{sec:regular}). 
\item An intrinsic radial component, $\mathbf{B}_{\textrm{rad}}$ (see Sec.~\ref{sec:radial}).
\item A random component, $\mathbf{B}_{\textrm{rdm}}$ (see Sec.~\ref{sec:random}).
\end{enumerate}

The magnetic field is defined at every point within the box by $\mathbf{B}_{\textrm{total}}=\mathbf{B}_{\textrm{reg}}+\mathbf{B}_{\textrm{rdm}}+\mathbf{B}_{\textrm{rad}}$, including the very centre of the box (i.e., the centre of the SNR). In the physical picture of an SNR, we expect a very small magnetic field at the centre (or it may be dominated by a compact object), but in practice this does not impact the modelling since we scale the CRE density such that it is zero in the centre of the SNR (see Sec.~\ref{sec:cre}) and thus the scale of the magnetic field at the centre is not relevant.

The coordinate system is defined such that $B_x$ is the line-of-sight component, $B_y$ is the horizontal component when viewed on a projected image, and $B_z$ is the vertical component. We do not explicitly enforce that the resultant magnetic field be divergence free since this should not impact our conclusions, which are based on comparing the effect of the different components.

\begin{figure}

\begin{flushleft}
\includegraphics[height=6cm]{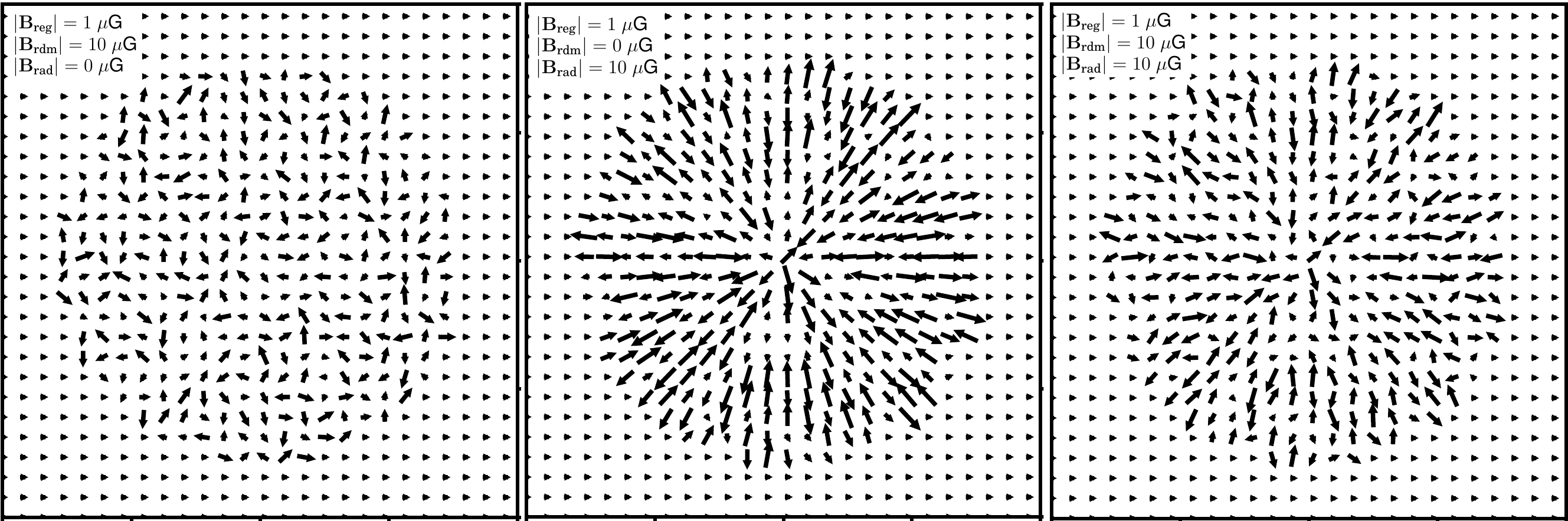}

\end{flushleft}

\caption{\label{fig:bfield}Magnetic field,   
$\mathbf{B}_{\textrm{total}}=\mathbf{B}_{\textrm{reg}}+\mathbf{B}_{\textrm{rdm}}+\mathbf{B}_{\textrm{rad}}$, shown for the three cases modeled in this study. We show the y-z plane cut through the centre of the SNR. 
}
\end{figure}

\begin{figure}

\begin{flushleft}
\includegraphics[height=5.4cm]{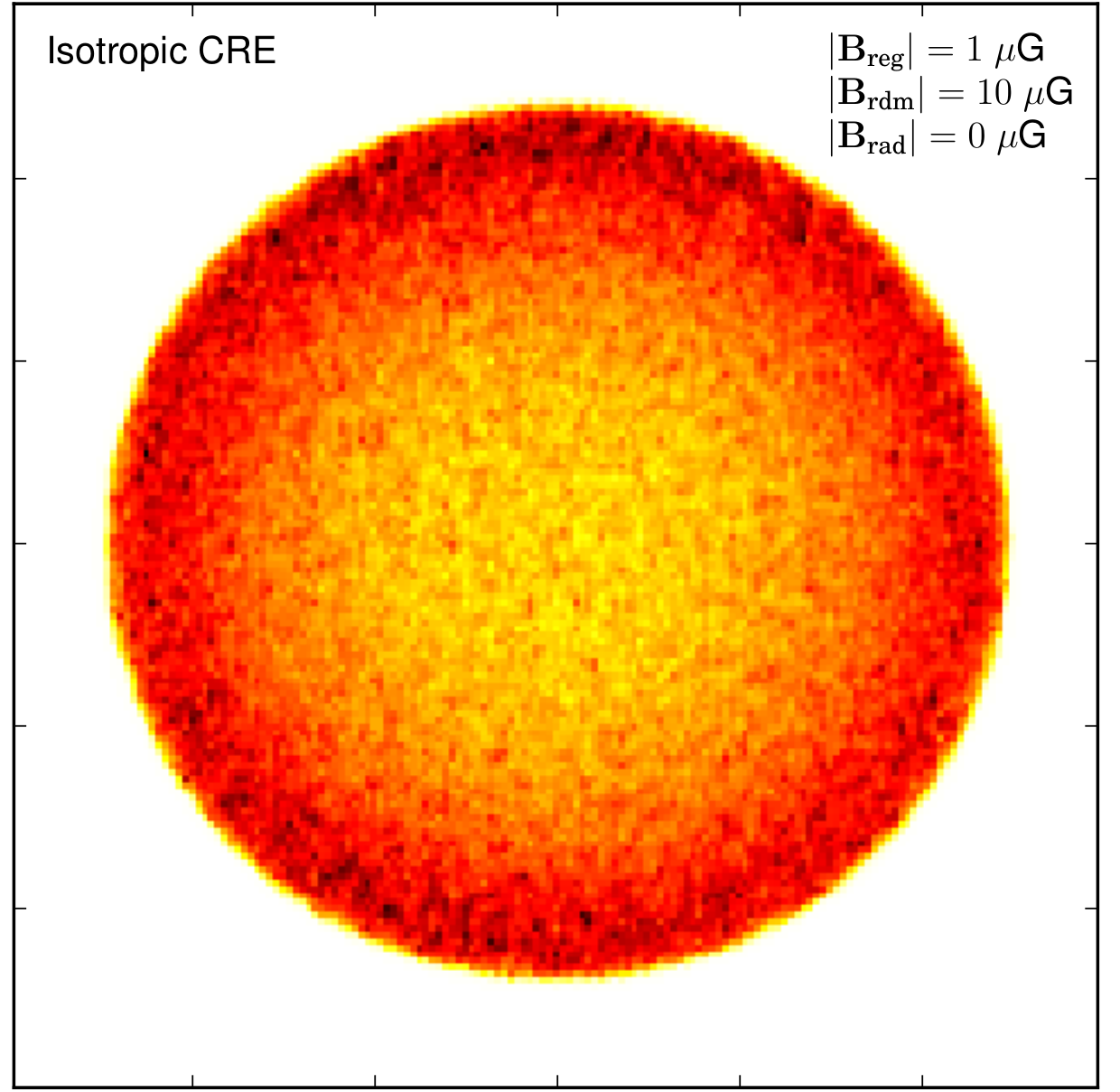}
\includegraphics[height=5.4cm]{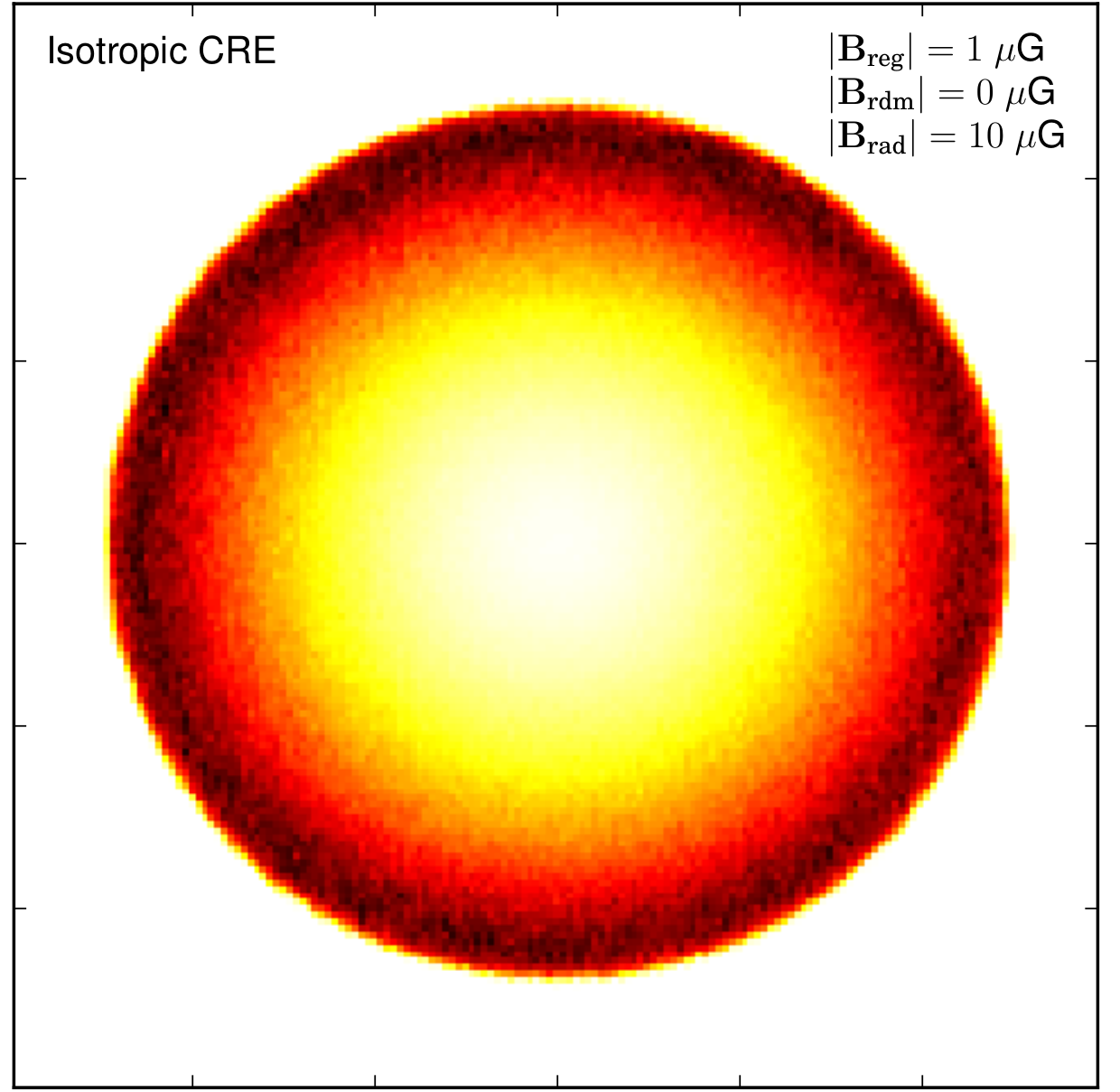}
\includegraphics[height=5.4cm]{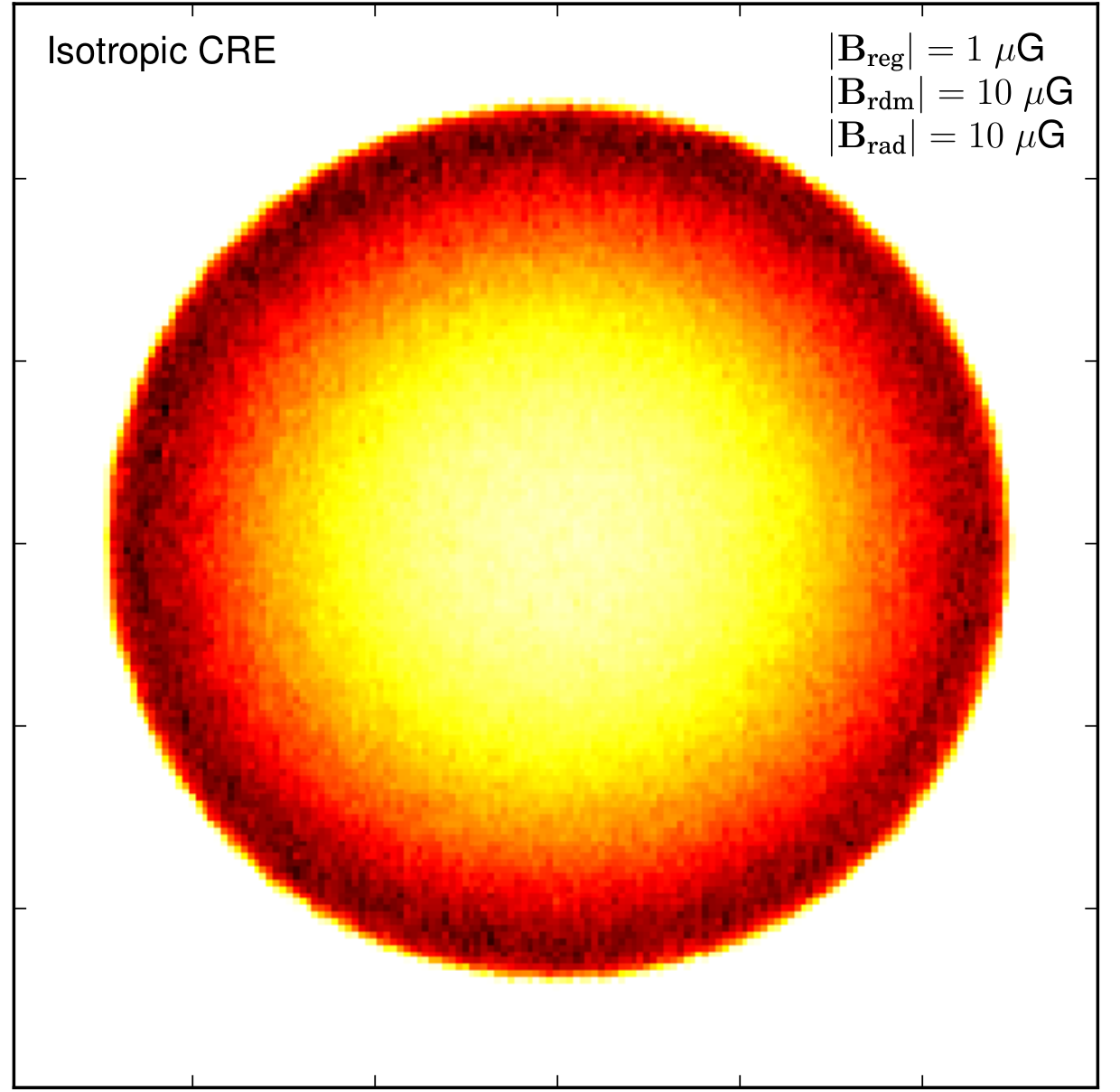}
\includegraphics[height=5.5cm]{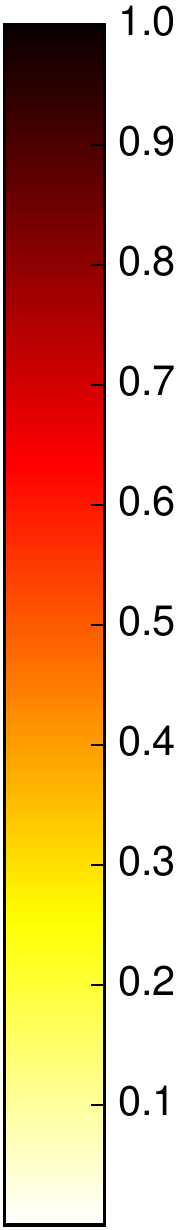}
\includegraphics[height=5.4cm]{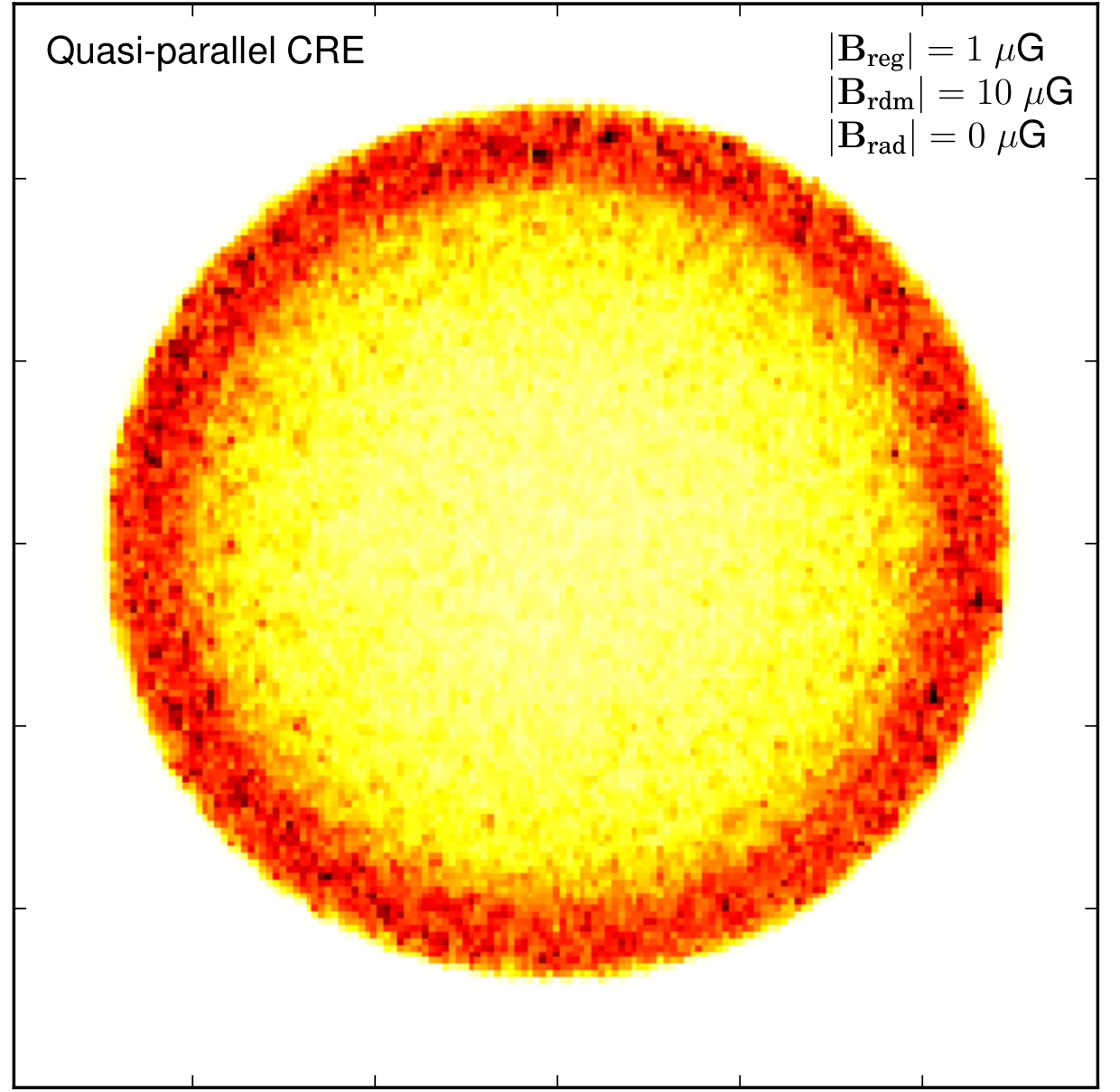}
\includegraphics[height=5.4cm]{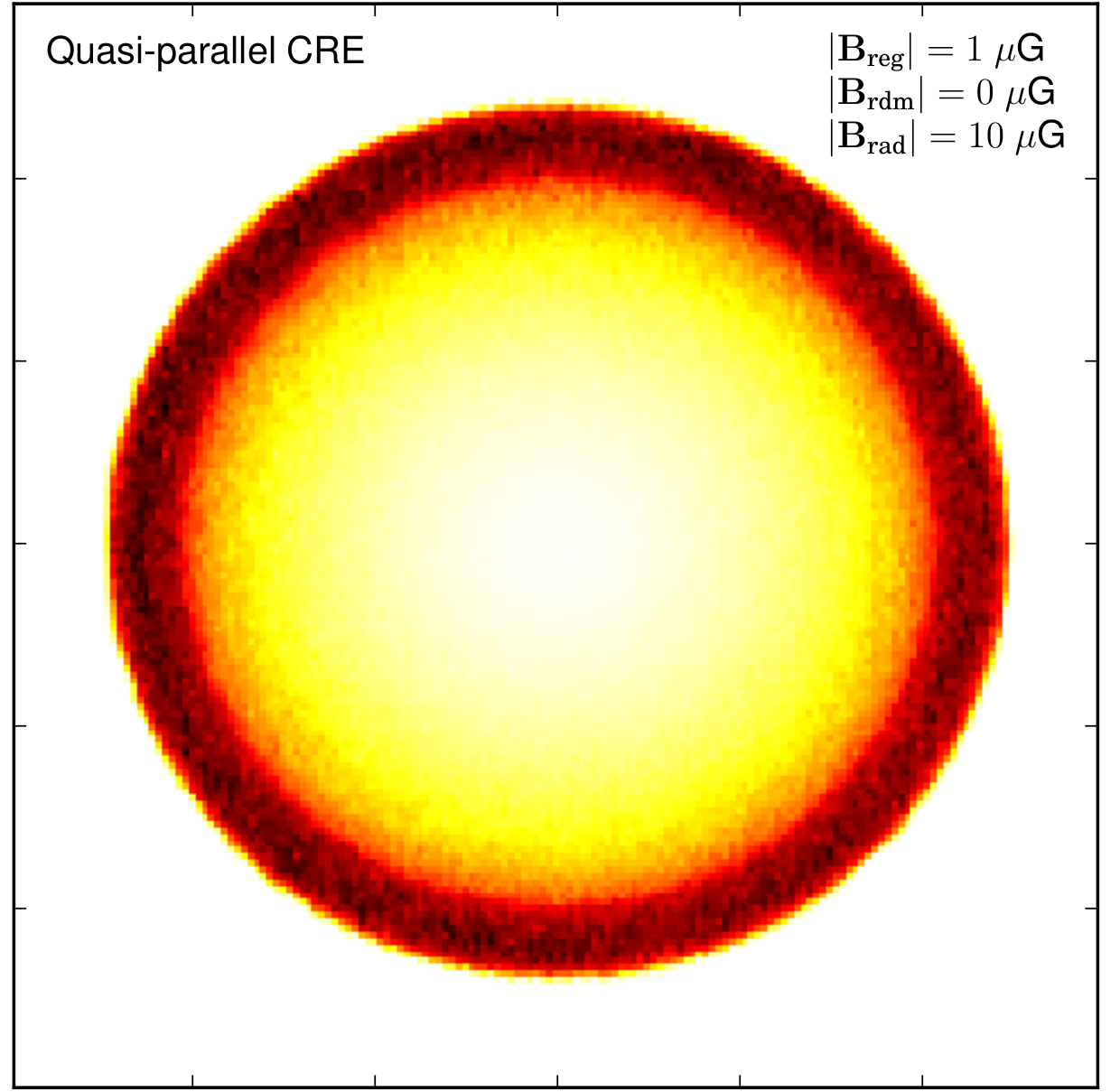}
\includegraphics[height=5.4cm]{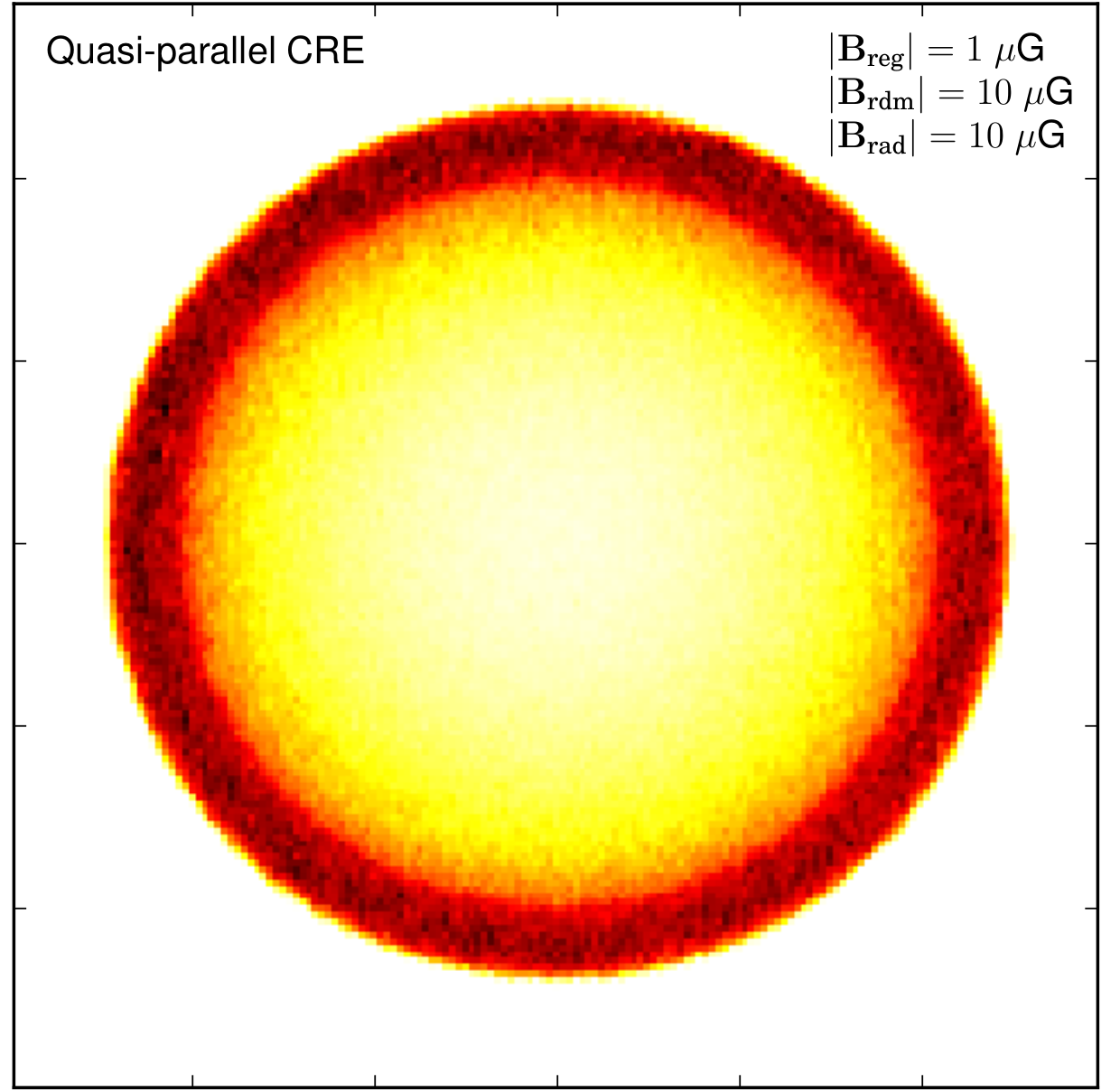}
\includegraphics[height=5.5cm]{f3.pdf}
\includegraphics[height=5.4cm]{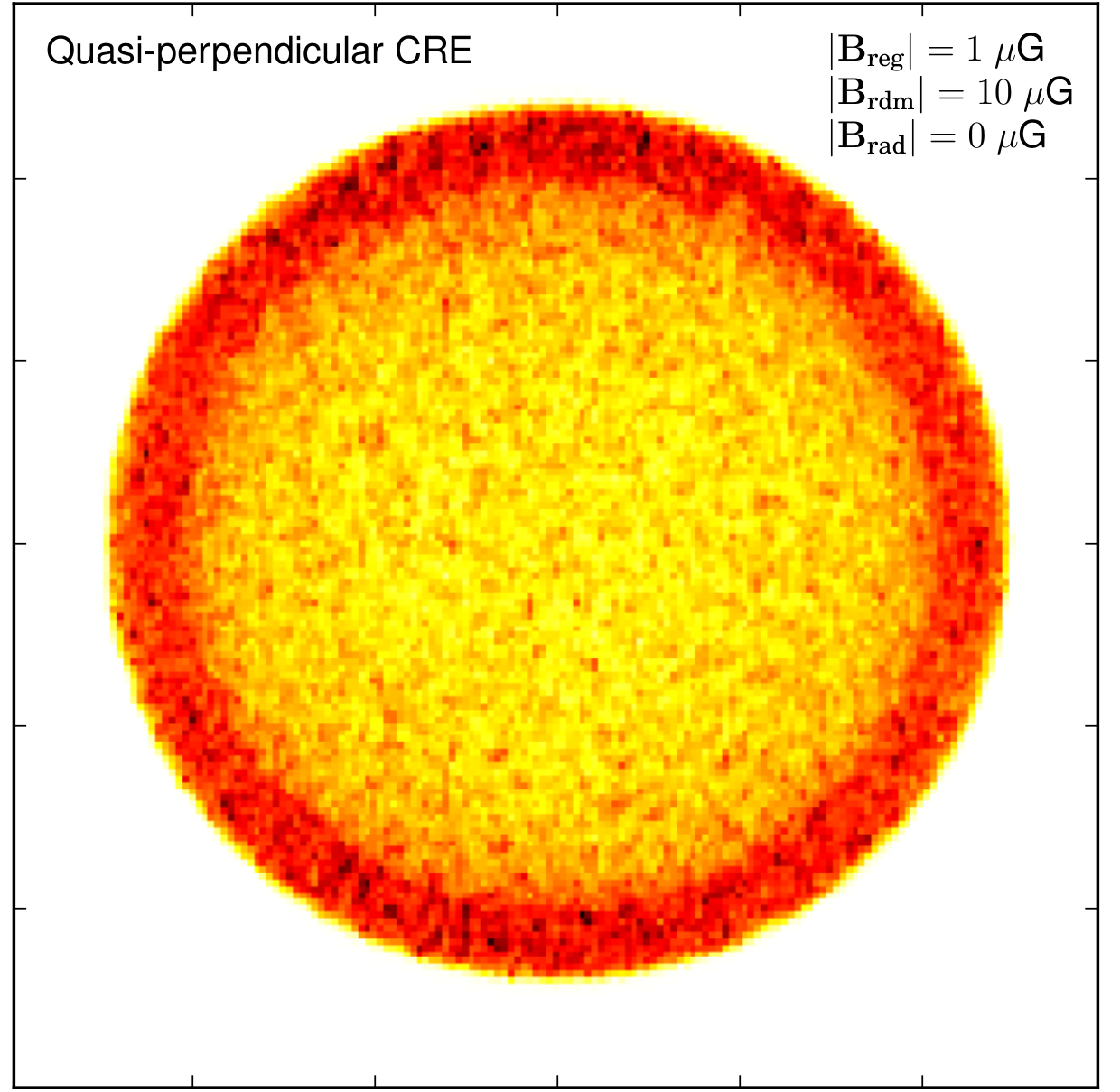}
\includegraphics[height=5.4cm]{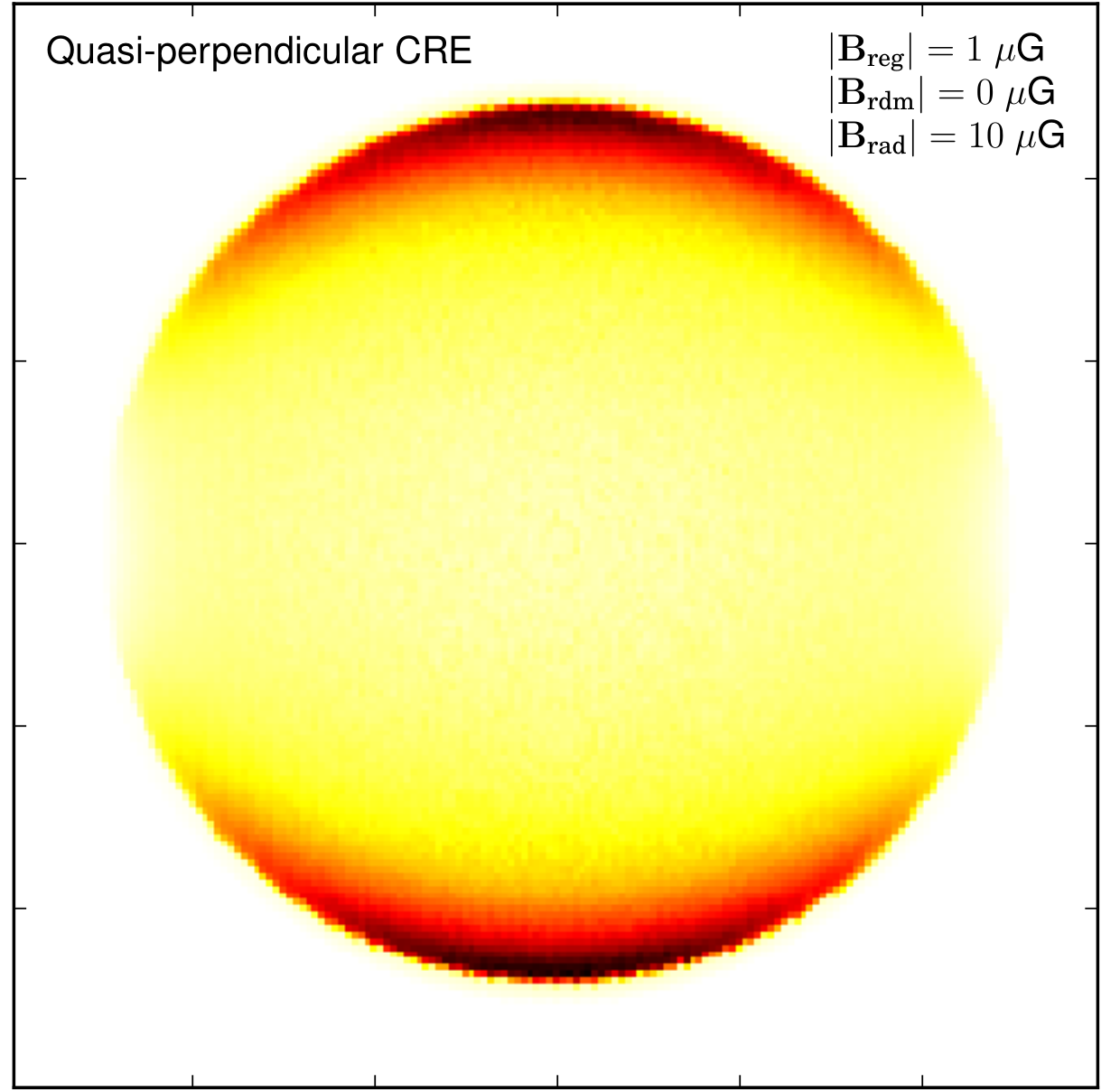}
\includegraphics[height=5.4cm]{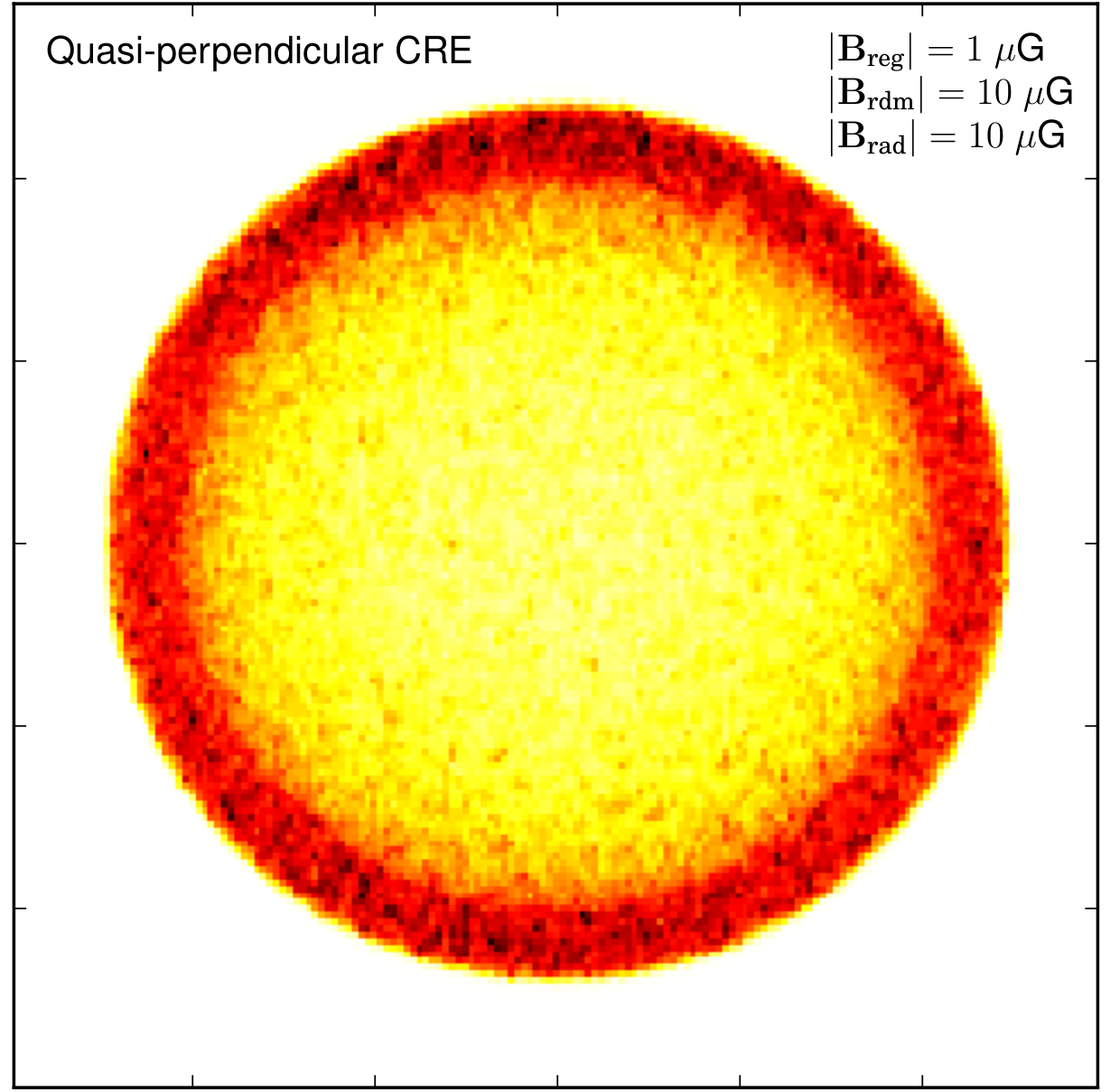}
\includegraphics[height=5.5cm]{f3.pdf}
\end{flushleft}

\caption{\label{fig:i}Simulated total intensity maps resulting from the magnetic field configurations illustrated in Fig.~\ref{fig:bfield} and shown with isotropic (top row), quasi-parallel (centre row), and quasi-perpendicular (bottom row) CRE acceleration mechanisms.} 
\end{figure}

\begin{figure}

\begin{flushleft}
\includegraphics[height=5.4cm]{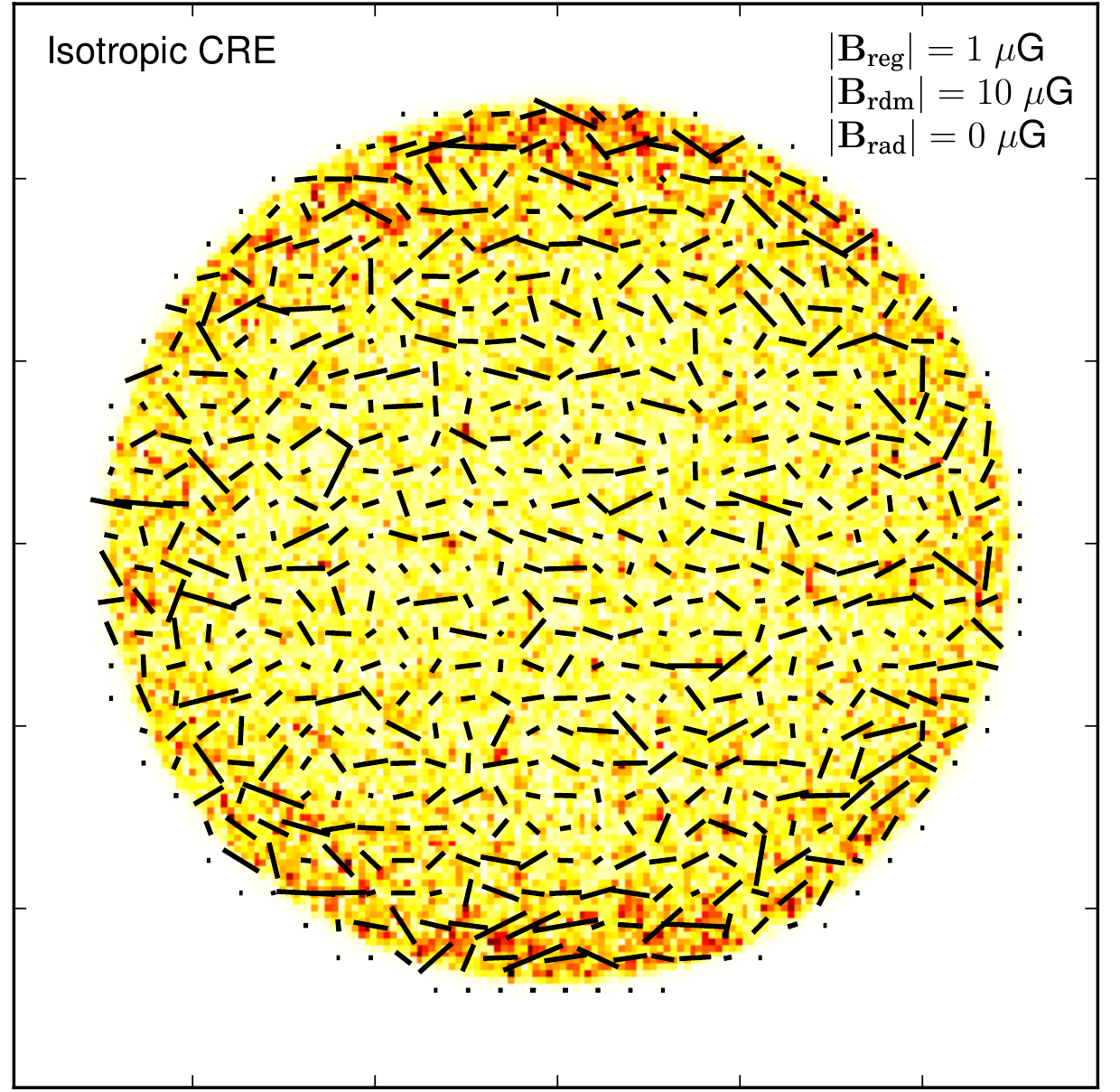}
\includegraphics[height=5.4cm]{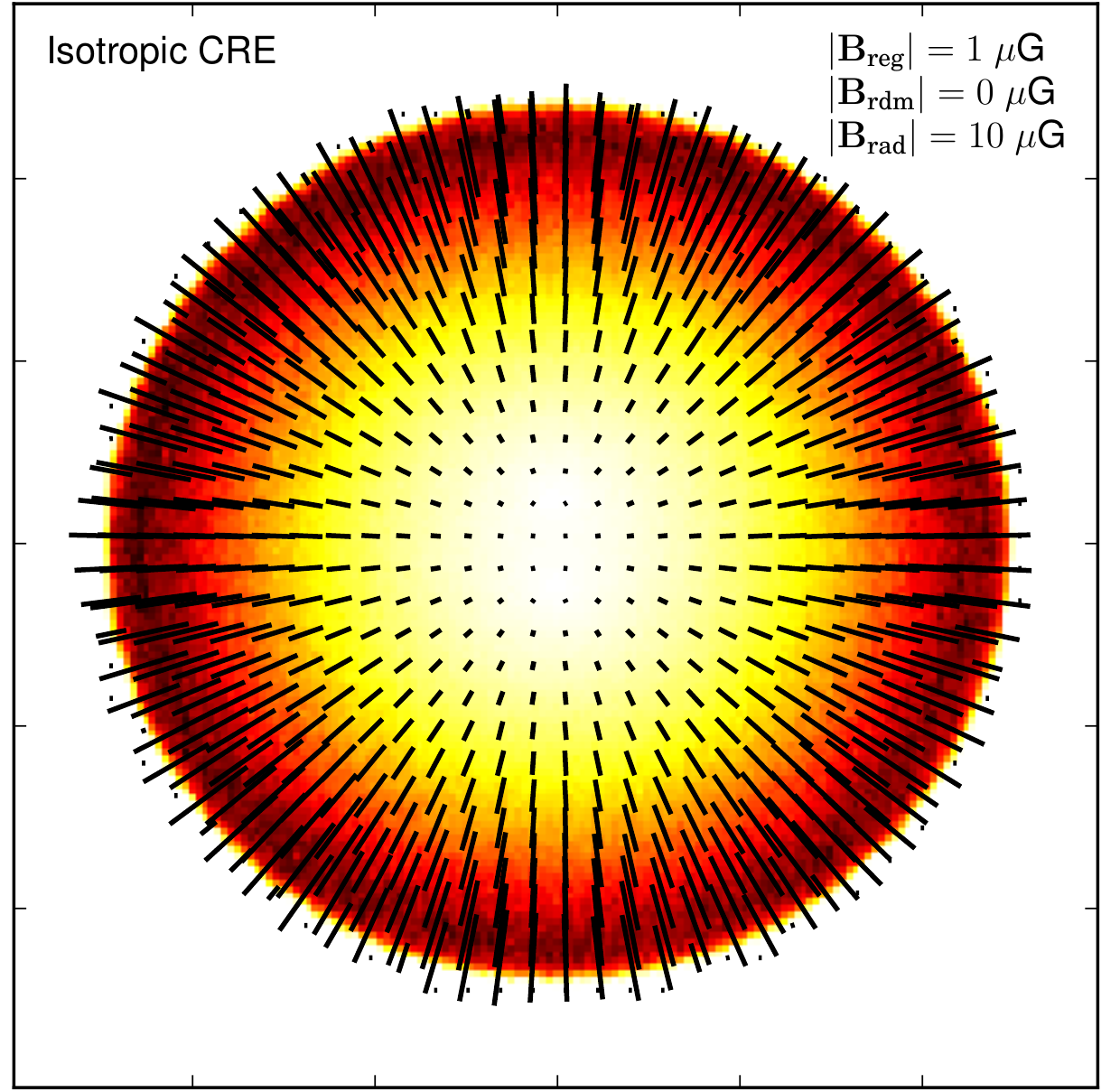}
\includegraphics[height=5.4cm]{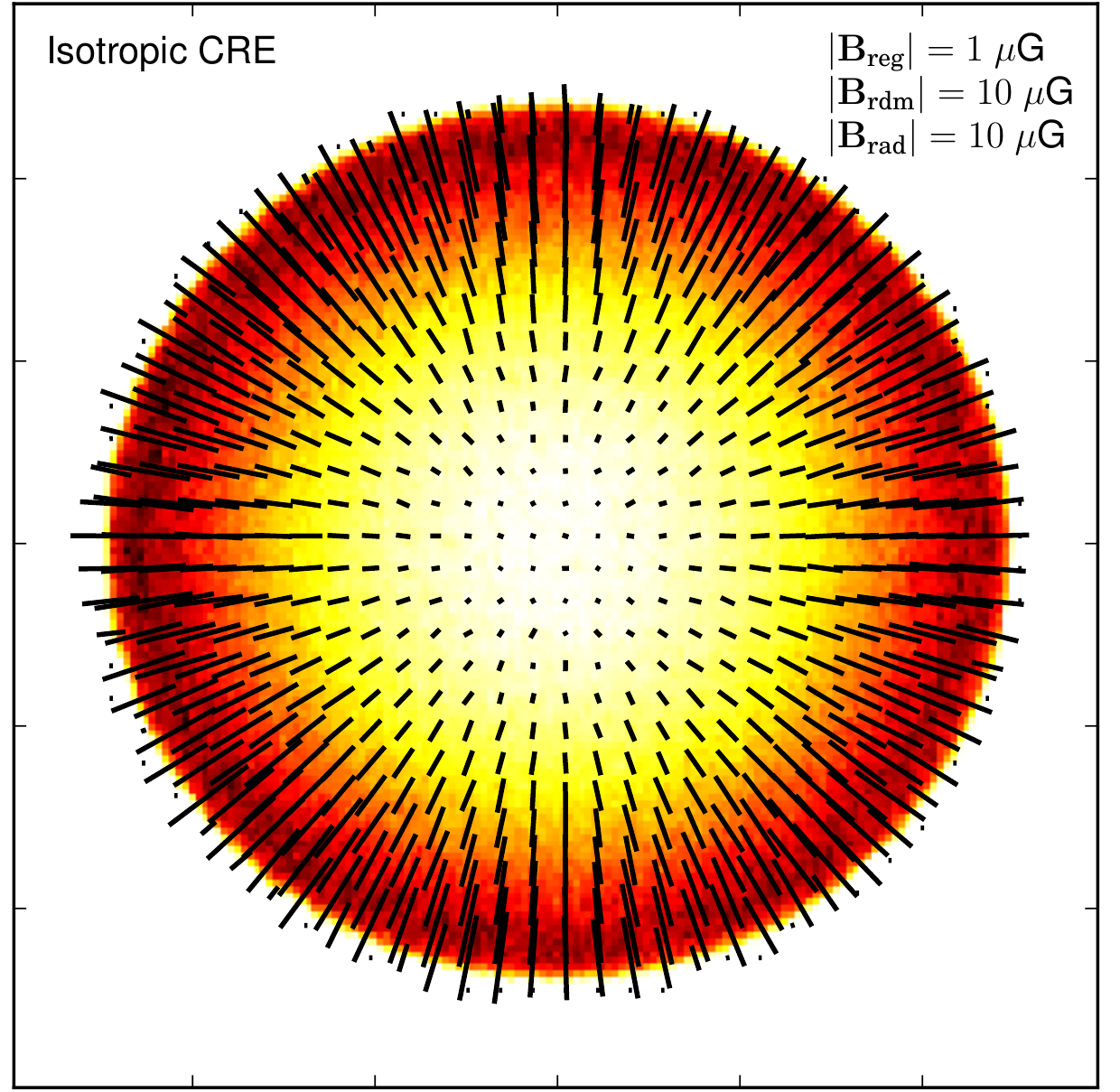}
\includegraphics[height=5.4cm]{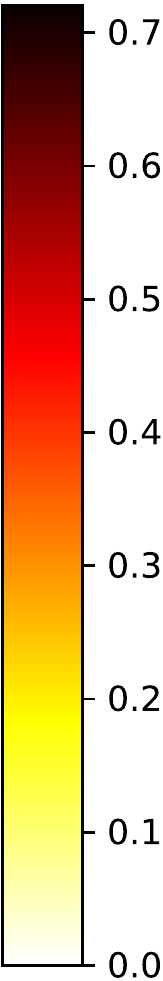}
\includegraphics[height=5.4cm]{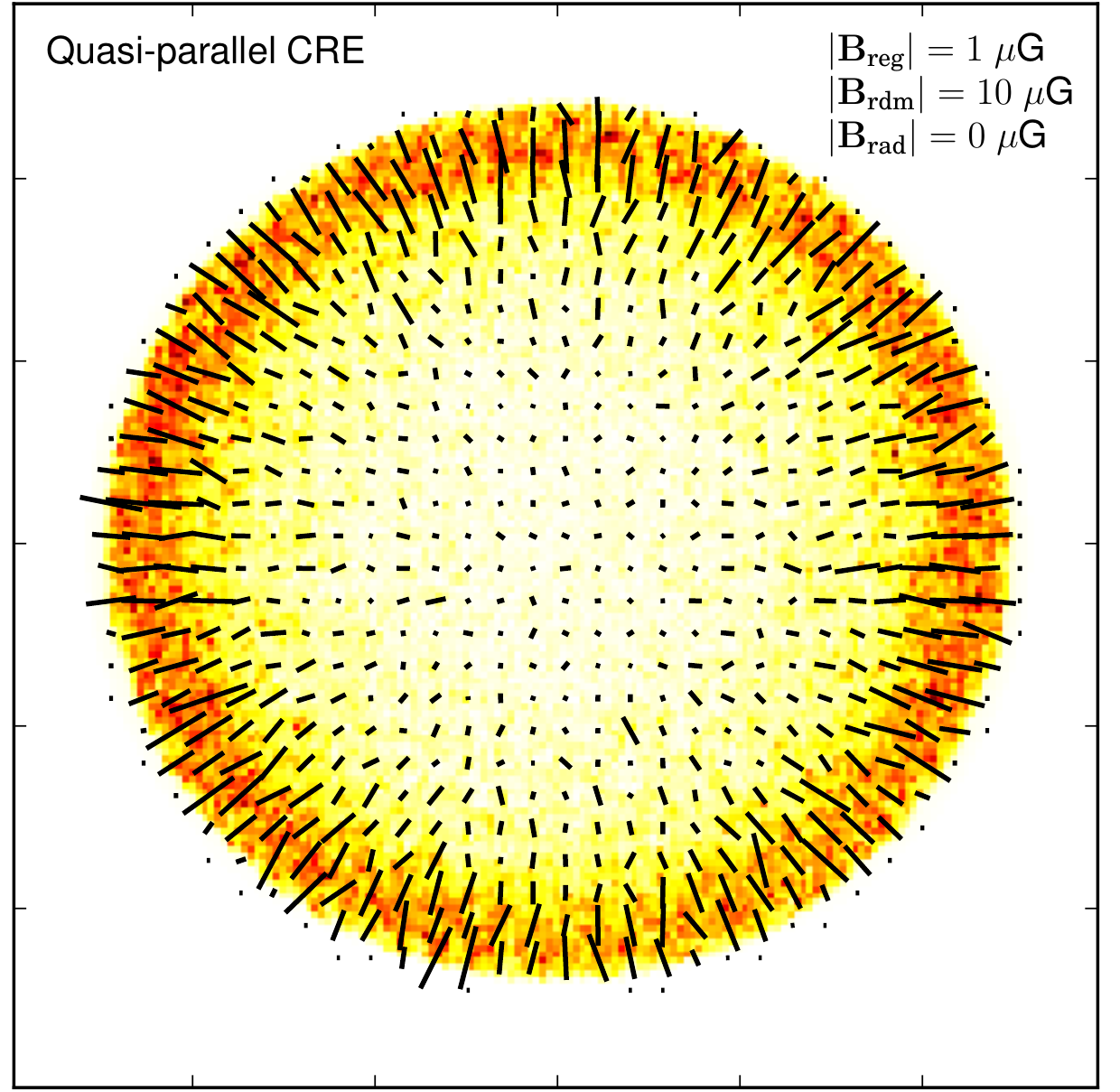}
\includegraphics[height=5.4cm]{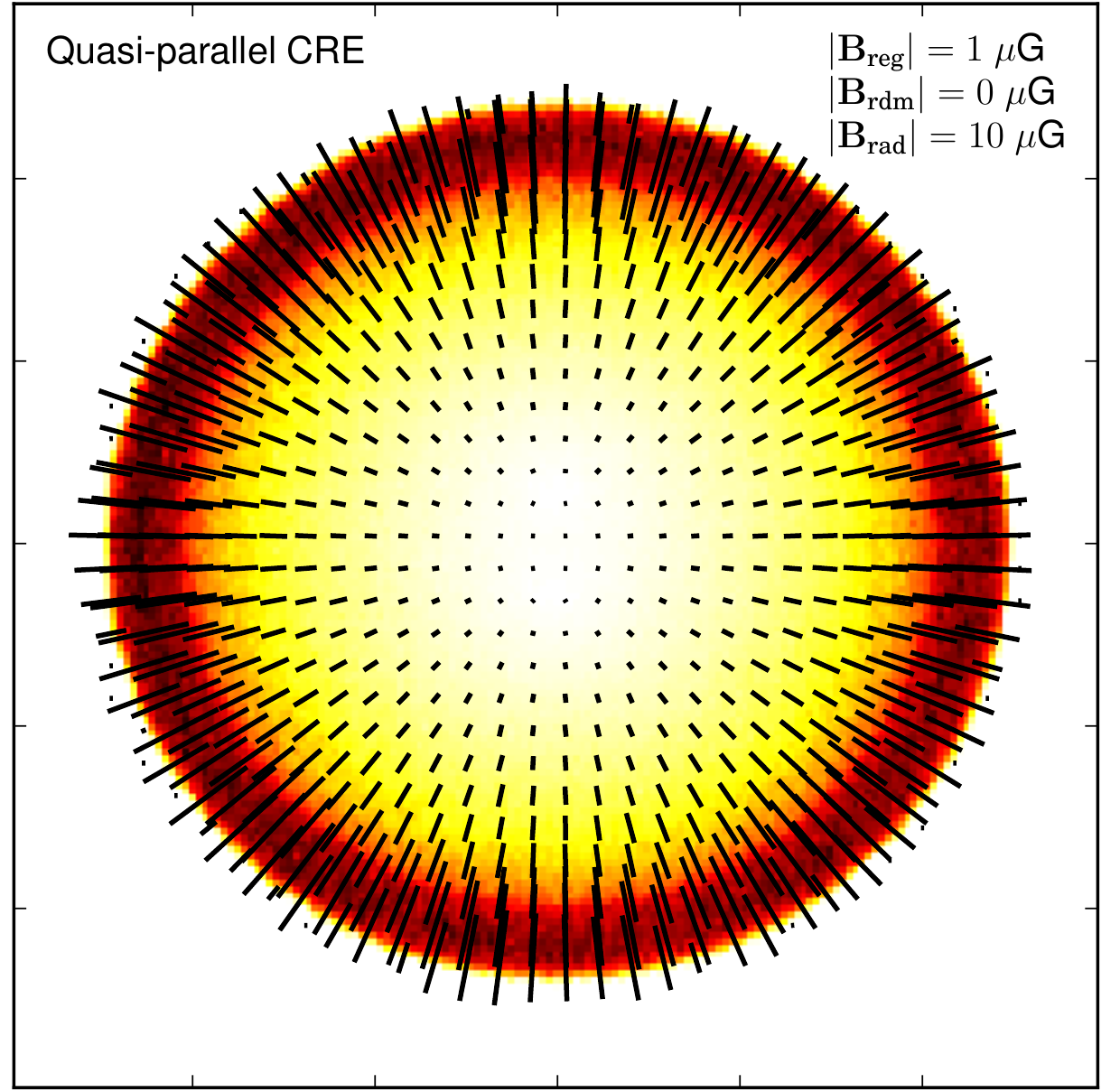}
\includegraphics[height=5.4cm]{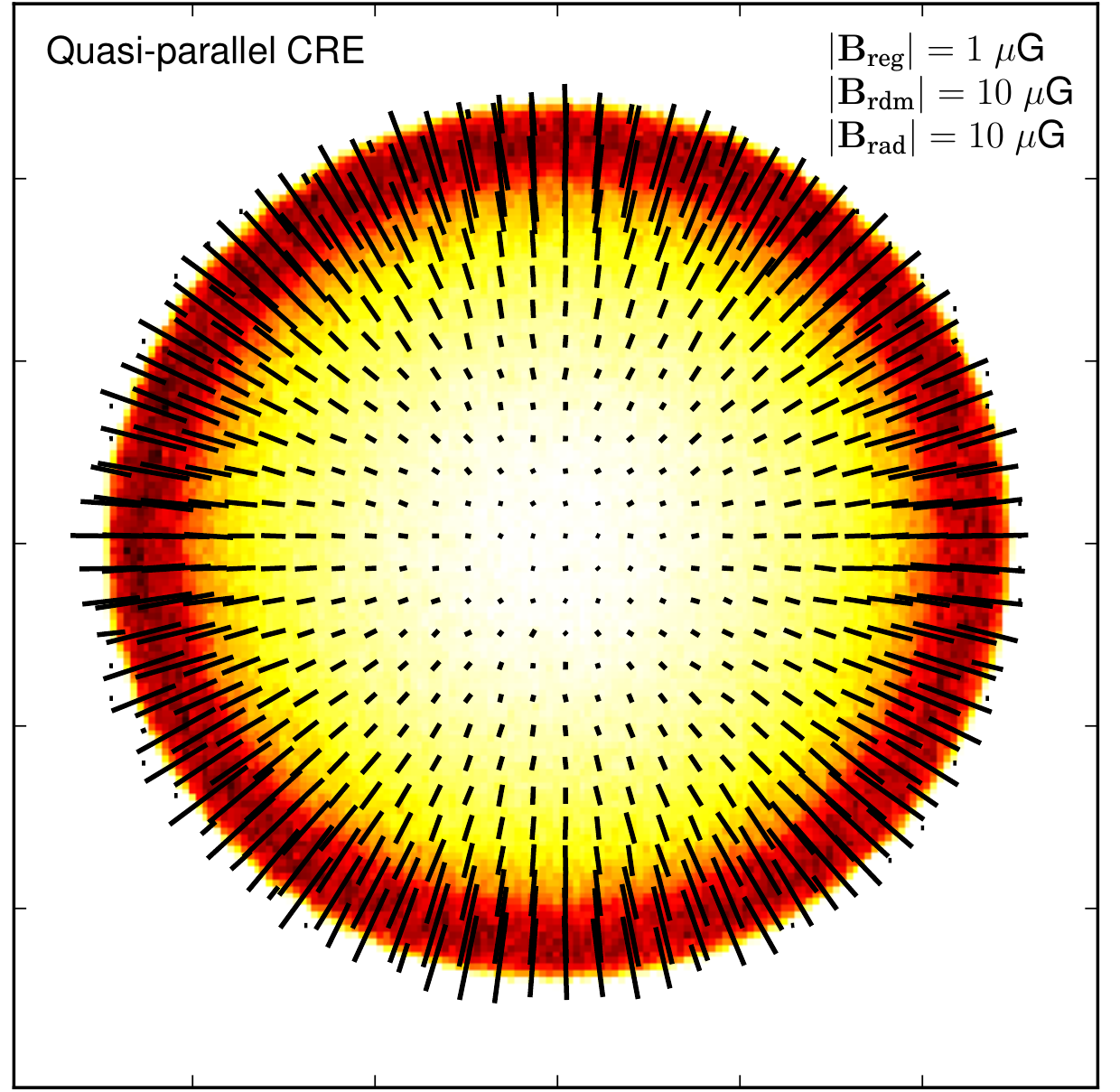}
\includegraphics[height=5.4cm]{f4.pdf}
\includegraphics[height=5.4cm]{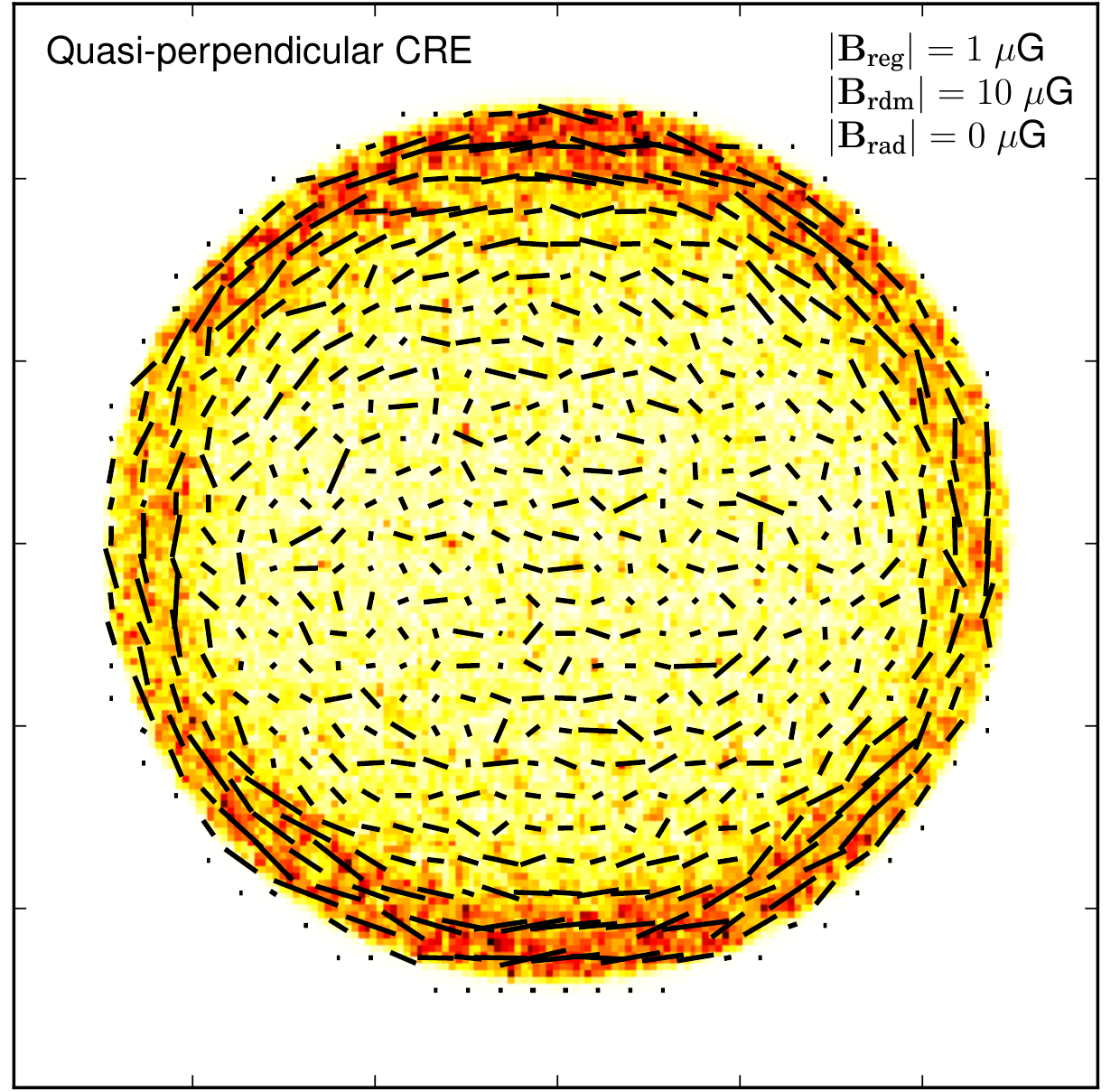}
\includegraphics[height=5.4cm]{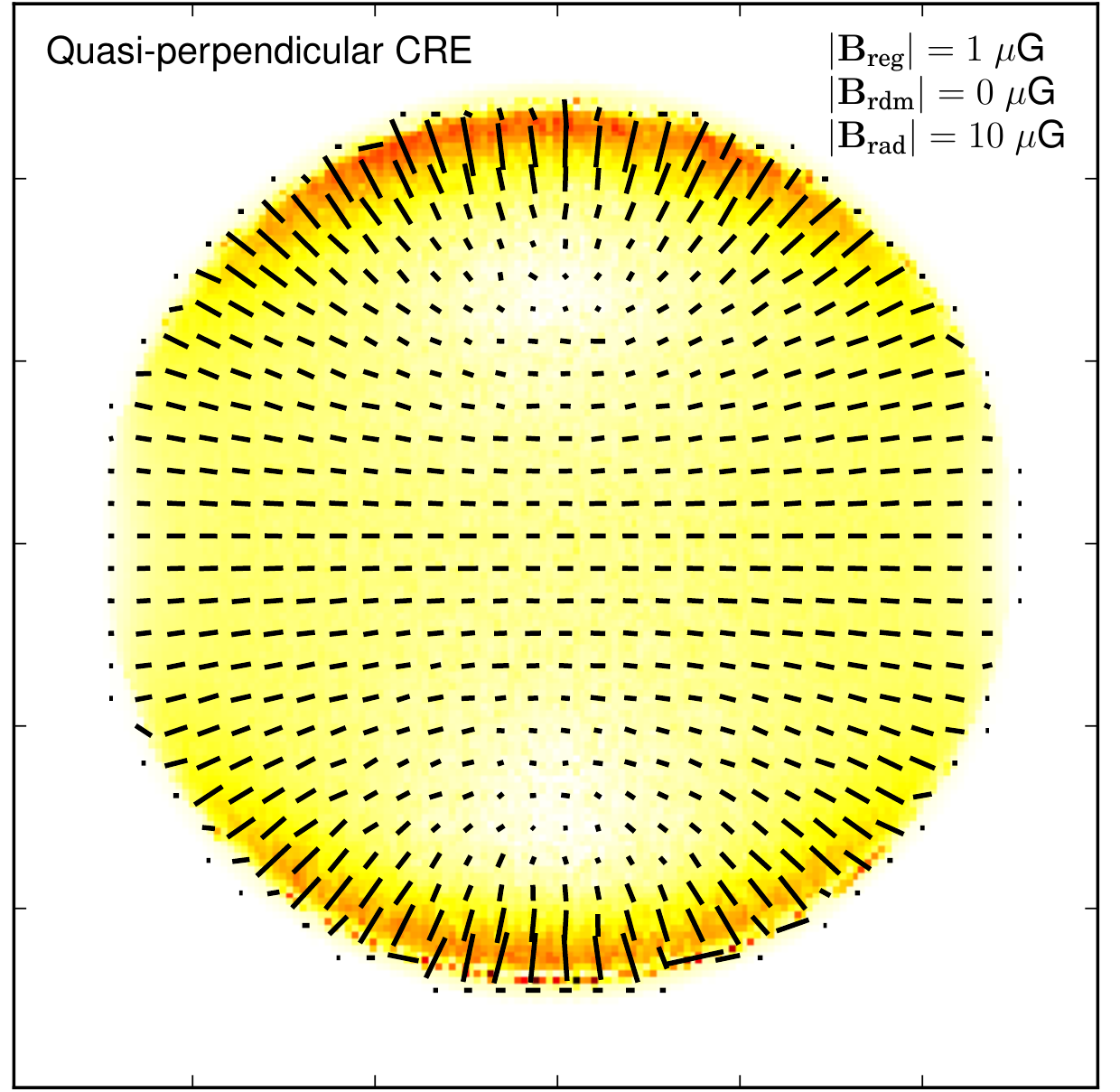}
\includegraphics[height=5.4cm]{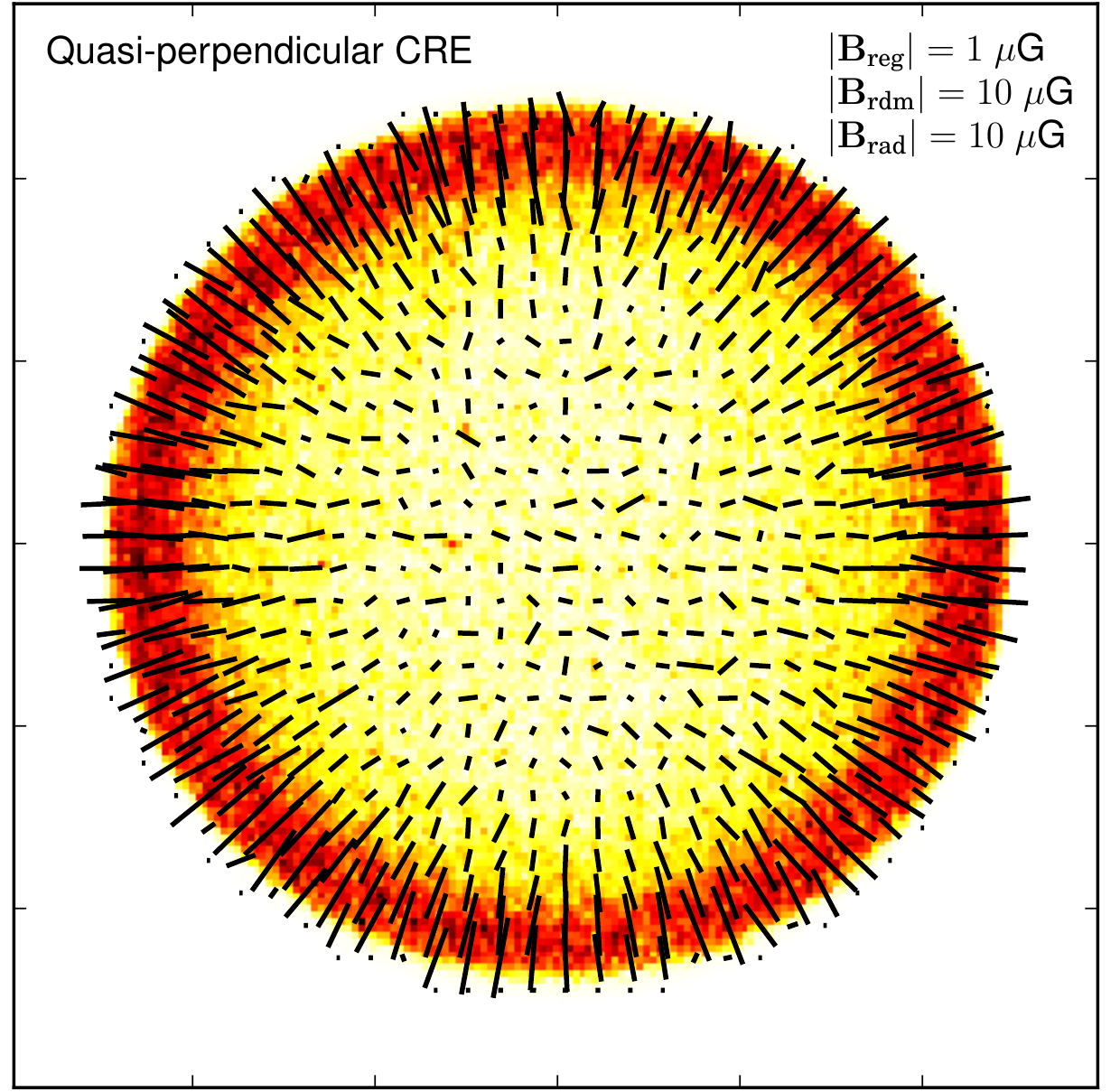}
\includegraphics[height=5.4cm]{f4.pdf}
\end{flushleft}
\caption{\label{fig:pi}Same cases as in Fig.~\ref{fig:i} but for simulated polarized intensity with the projected magnetic field vectors plotted using the simulated observations.}\end{figure}

\begin{figure}

\begin{flushleft}

\includegraphics[width=5.9cm]{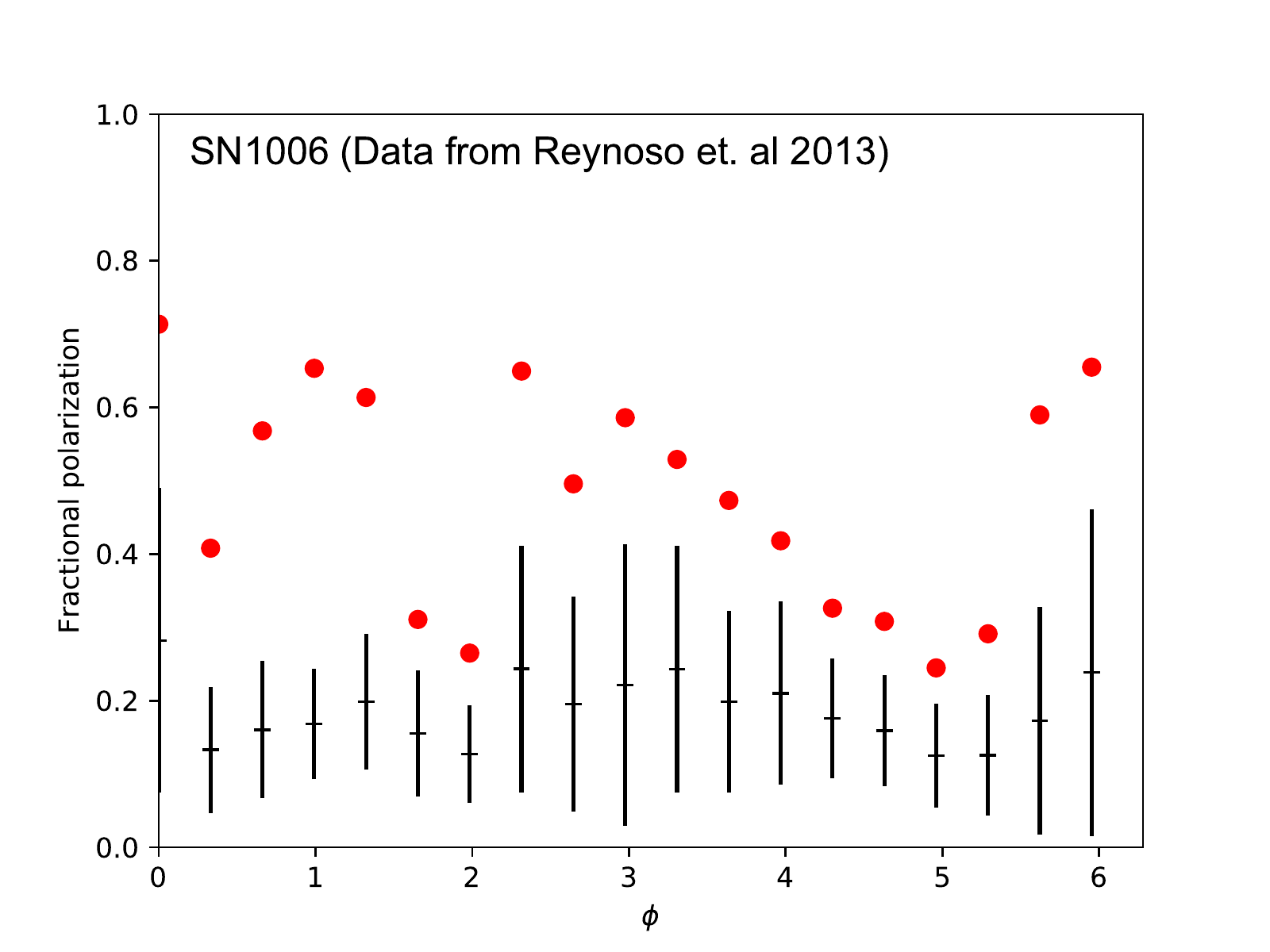}
\includegraphics[width=5.9cm]{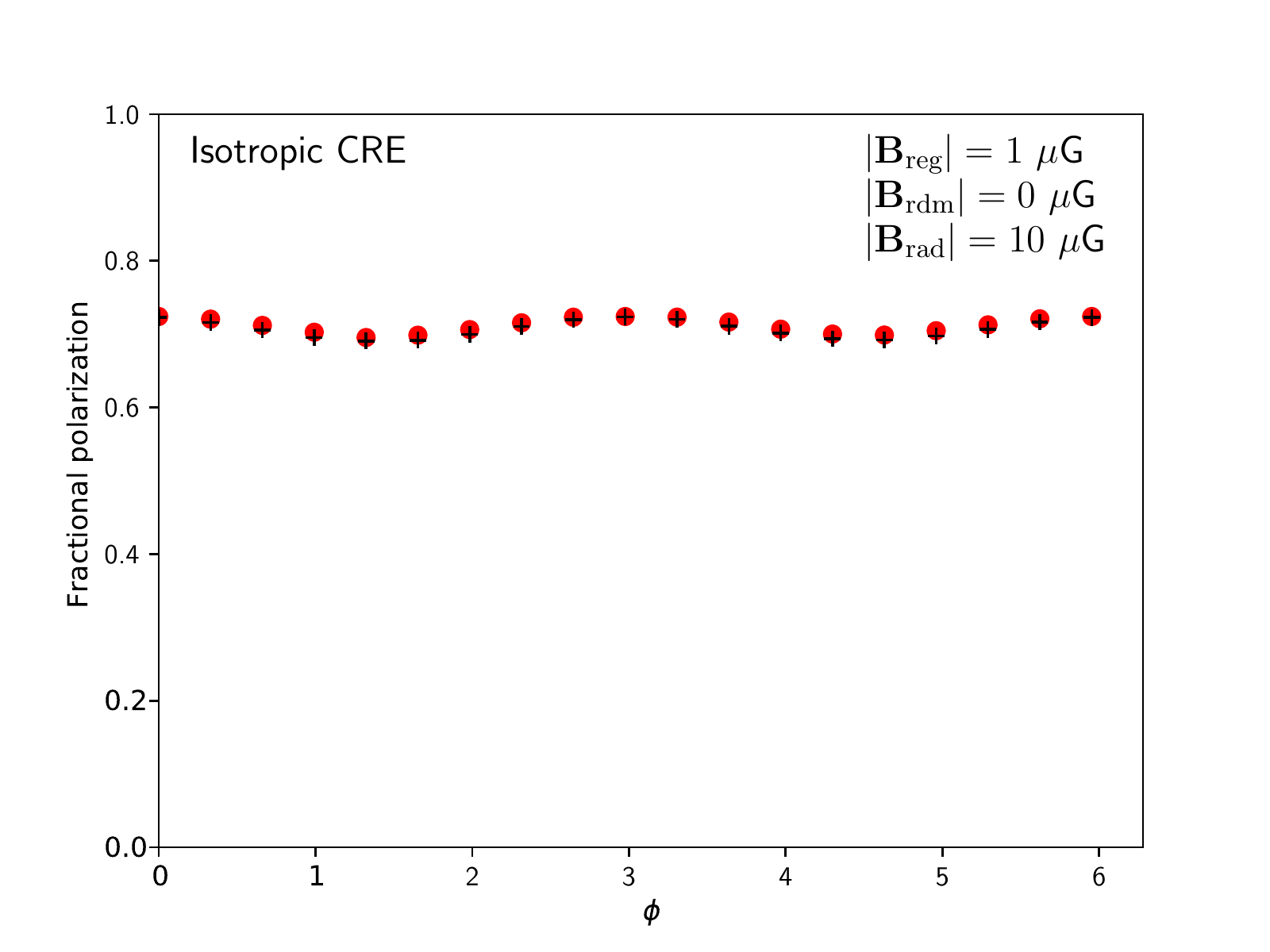}
\includegraphics[width=5.9cm]{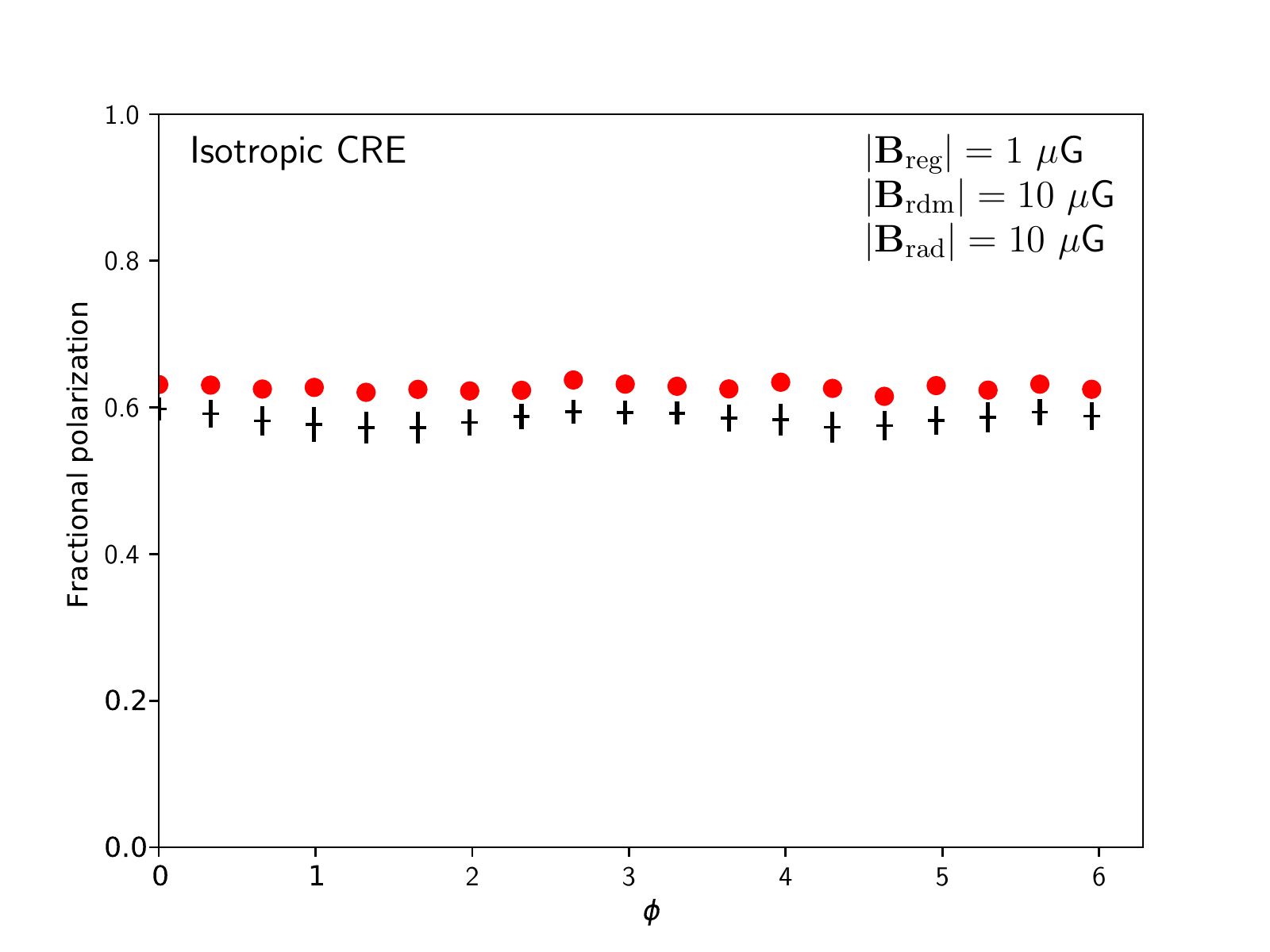}
\includegraphics[width=5.9cm]{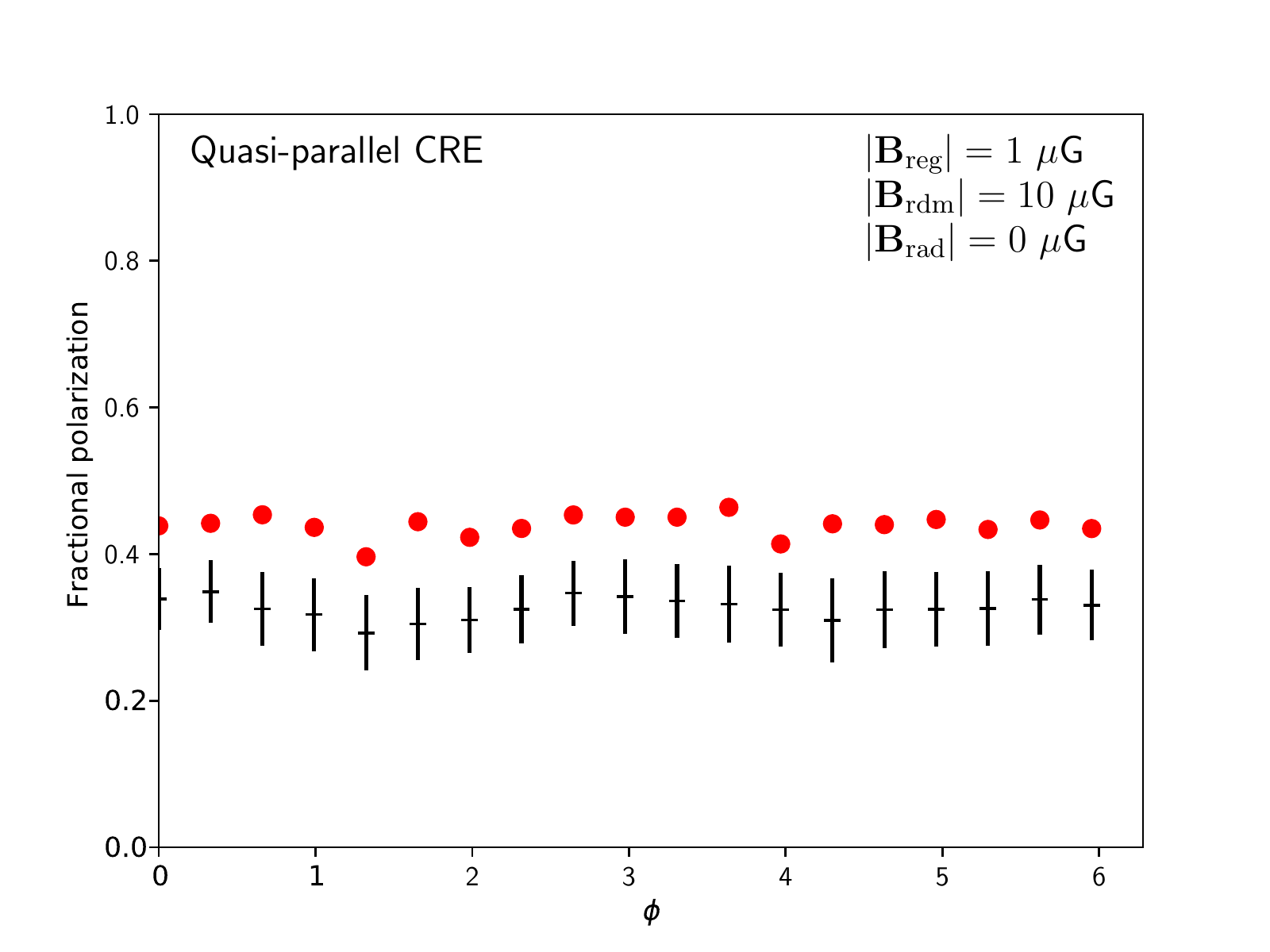}
\includegraphics[width=5.9cm]{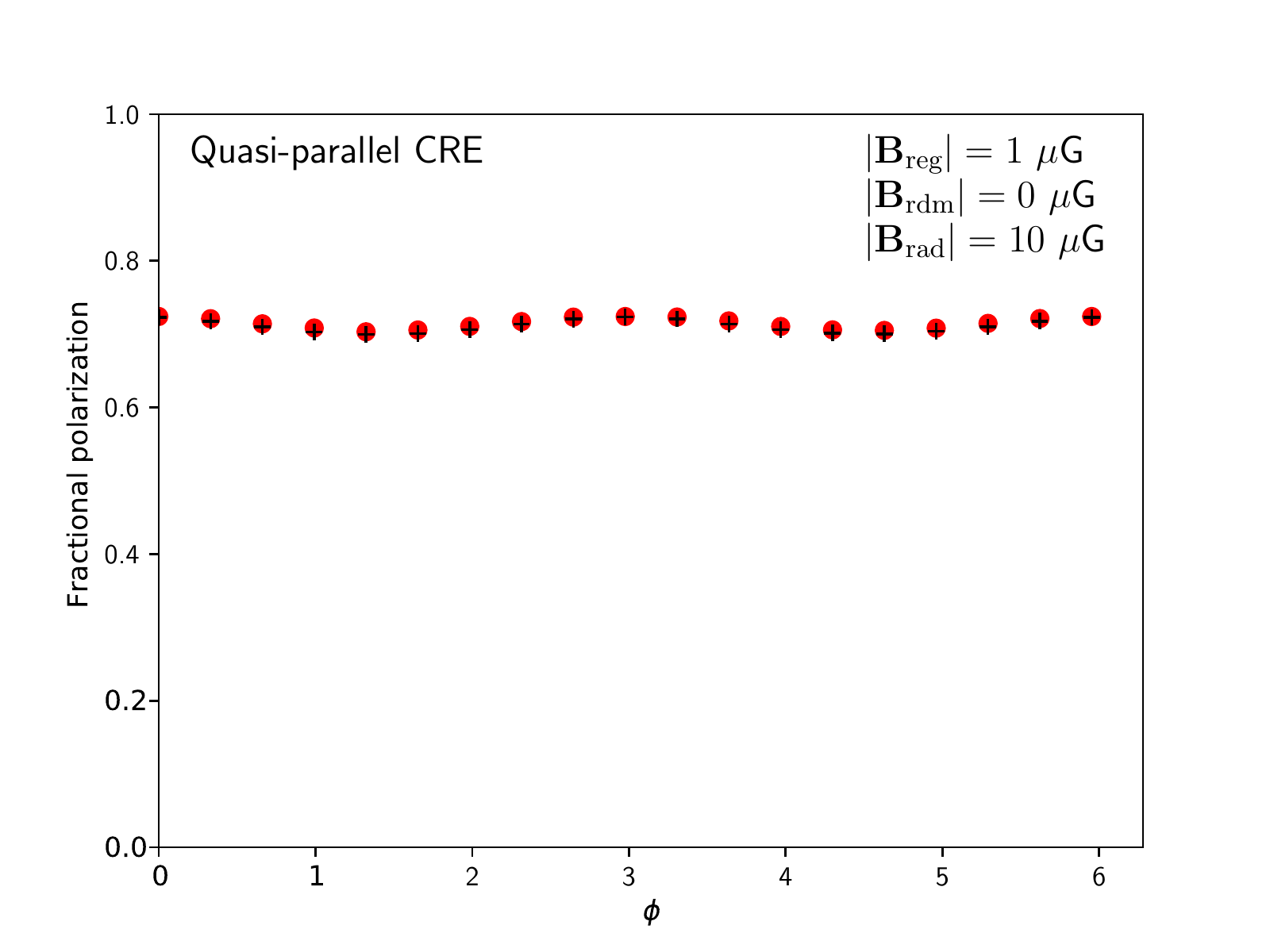}
\includegraphics[width=5.9cm]{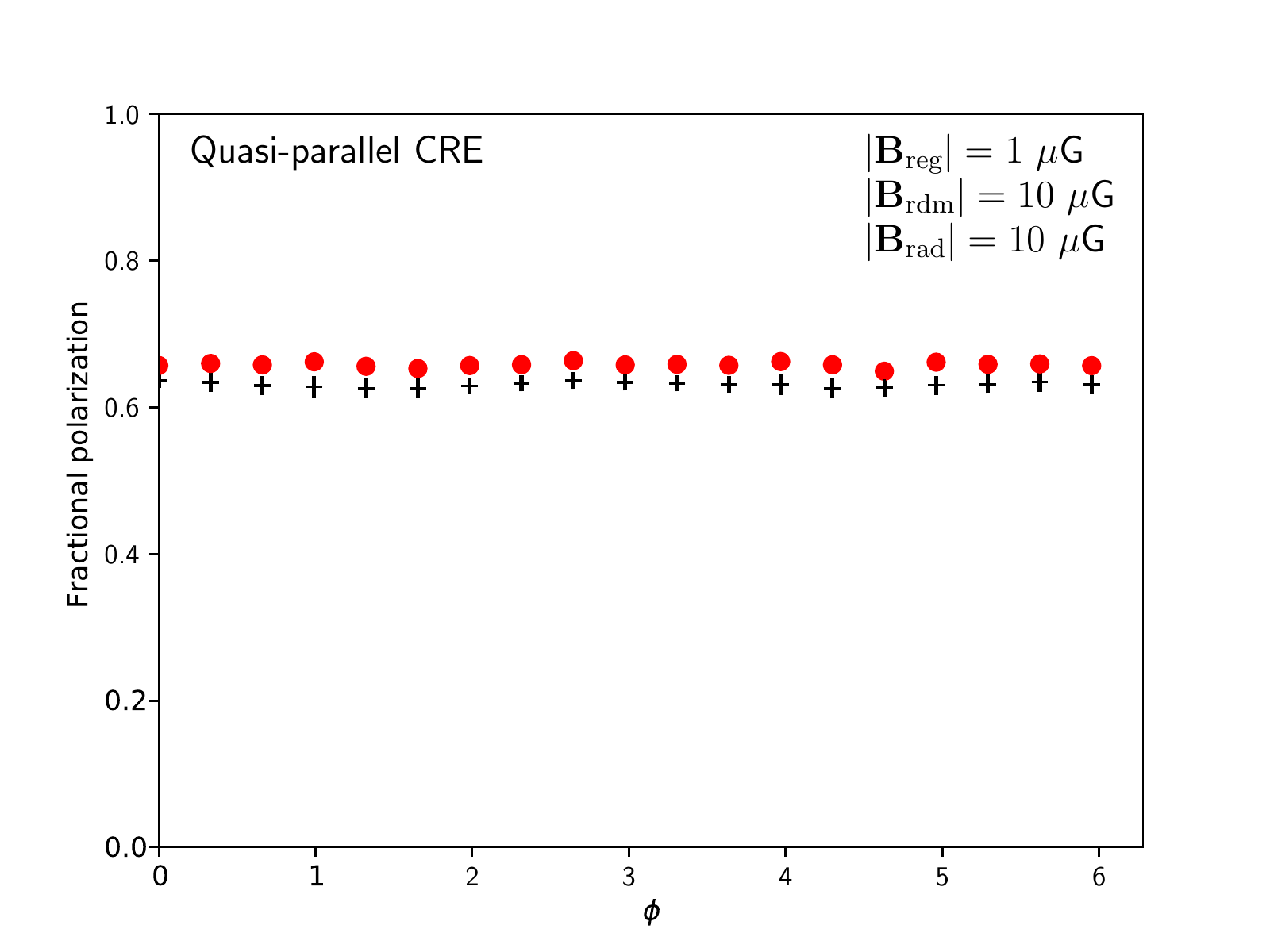}
\includegraphics[width=5.9cm]{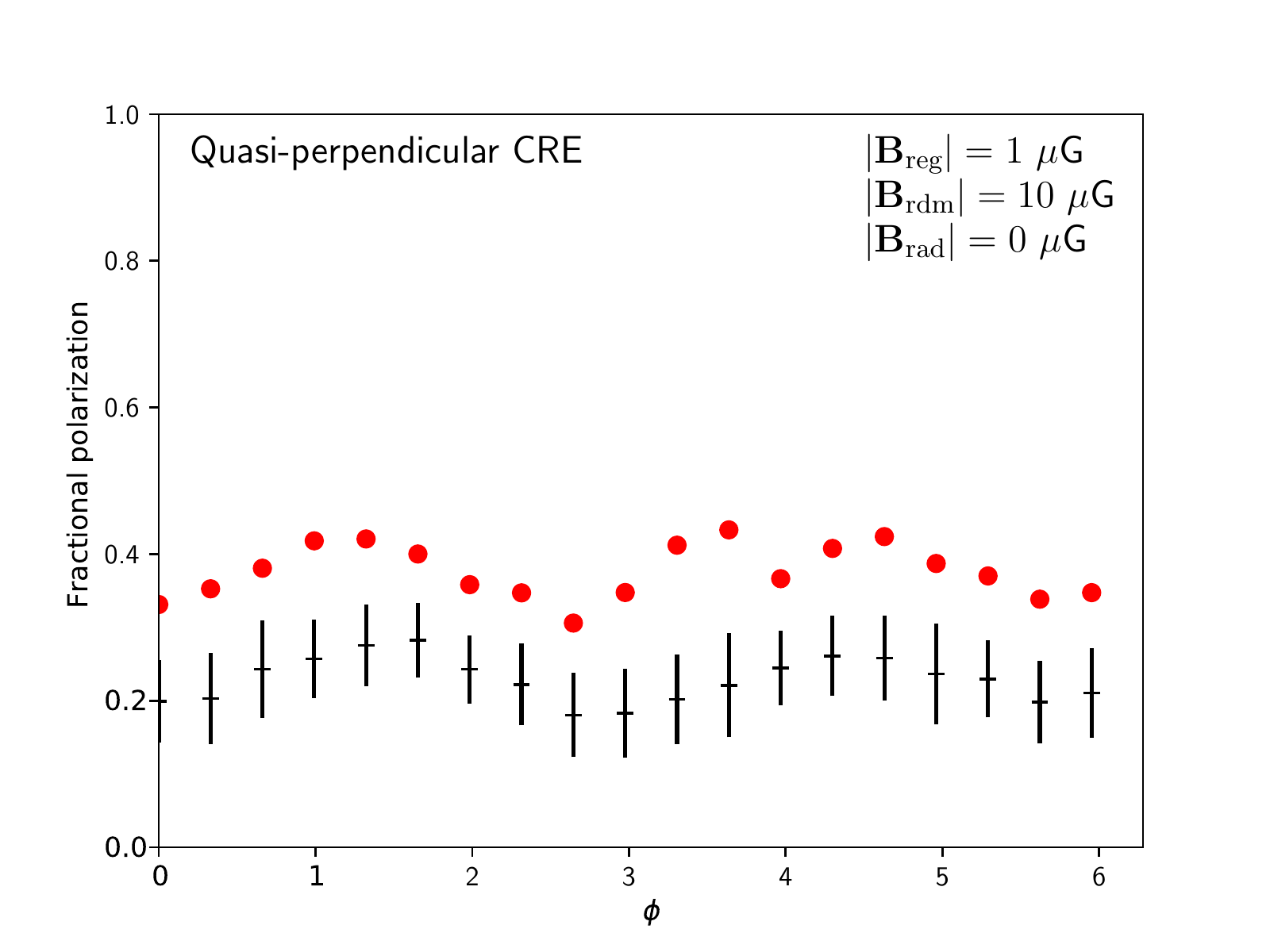}
\includegraphics[width=5.9cm]{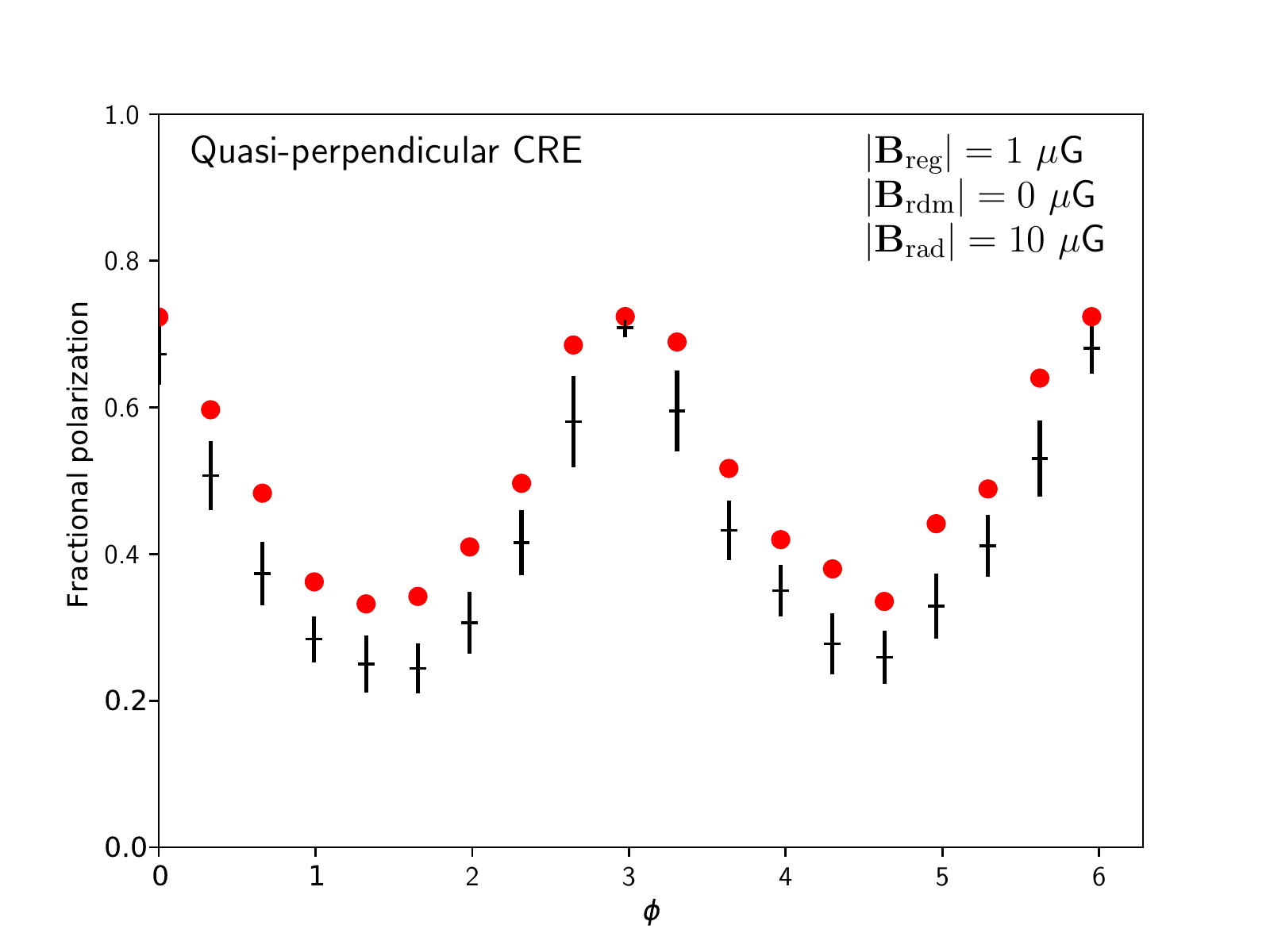}
\includegraphics[width=5.9cm]{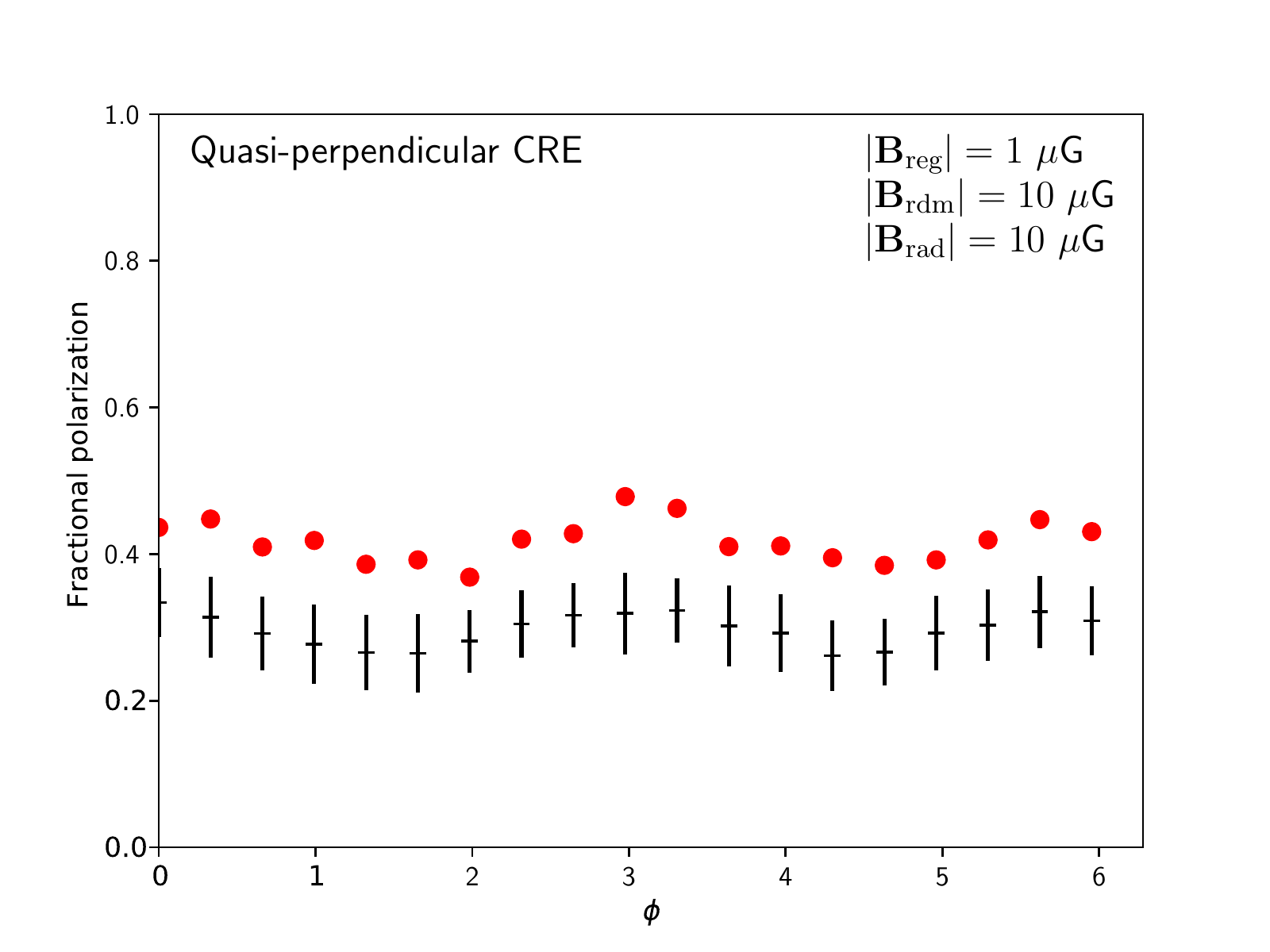}

\end{flushleft}
\caption{\label{fig:frac-plot} Plots of the fractional polarization vs. $\phi$. We define an annulus around the SNR with an inner radius set to 90\% of the total radius and show the mean (black) and maximum (red) values measured in 20 azimuthal bins. We show the same simulated cases as Fig.~\ref{fig:i}, with the exception of the top-left case, which has been replaced by a plot of SN1006 data from \citet{Reynoso:2013tr}. The data have similarities to the radially dominated quasi-perpendicular case (bottom-centre), however, we note that we have not attempted to make a best fit to SN1006.}
\end{figure}

\subsection{\label{sec:regular}Compressed ordered component}

We assume that the explosion takes place in an environment where there exists an ambient, ordered magnetic field that will be compressed by the shock wave according to a Sedov-Taylor explosion model \citep{1959sdmm.book.....S} by the method described by \citet{2016A&A...587A.148W}. This is a reasonable assumption since the SNRs in question (Kepler, Tycho, Cas A, and SN1006) are likely to be at least in transition to the Sedov phase. The Sedov mass density profile is a reasonable approximation even in a case where the transition has not fully taken place since we need only reasonable compression in the shock.

For these tests this field is initially oriented entirely parallel to the Galactic plane , with no line-of-sight component (i.e., $B_x$=0$~\mu$G, $B_y=1~\mu$G, $B_z$=0$~\mu$G).

\subsection{\label{sec:radial}Intrinsic radial component}
This component is constructed with an orientation that is parallel to the radius vector at all locations. Each point has a random amplitude between 0 and 1~$\mu$G and a random sign (i.e., the vector is either pointing towards the centre or towards the shock). The average amplitude is then normalized and scaled to a value appropriate for the particular model. The radial component is defined inside the shell radius, and is set to zero for all larger radii.

\subsection{\label{sec:random}Random component}

Using Hammurabi, we construct a Gaussian random-field given the rms amplitude (integrated over all scales), which we call $\sigma_{rdm}$, and the magnetic field power spectrum index, $\alpha$. The 3D power spectrum, $P(k)$, is given by $P(k)=k^{\alpha}$ where $k$ is the wavevector defined by $k=\sqrt({k_x^2+k_y^2+k_z^2})$. The 3D power spectrum is related to the energy spectrum, $E(k)$, which is given by $E(k)\propto k^2 P(k)$ \citep{2004ARA&A..42..211E}. For the case of a Kolmogorov spectrum, $E(k)\propto k^{-5/3}$, and thus $\alpha=-11/3$ and for the case of a flat energy spectrum, $E(k)\propto k^0$, or $\alpha=-2$.

We construct boxes with values of the magnetic field power spectrum index $\alpha=-2.5$ ($E(k)\propto k^{-1/2}$), which is nearly flat and consistent with the values predicted by \citet{2014ApJ...794...46C} from diffusive shock acceleration\footnote{See their Fig.~8, which shows that the slope of the power spectrum varies from negative to positive depending on the shock obliquity, and their footnote 2, p.8, which indicates that the power spectrum is flat in the relativistic regime}.  We use an outer turbulence scale of 1~pc. The inner scale is constrained by the pixel resolution (i.e., 0.04~pc). 

It should be noted that the simulations of \citet{2014ApJ...794...46C} are on much smaller scales than our models, in both space and time, but it is expected that turbulence will be excited up to the scales of the highest-energy particles, which are not included in these simulations. We use these along with results for turbulence in the Milky Way as a whole, which shows that the power spectrum is flatter than the Kolmogorov case \citep[e.g., ][]{2015ASSL..407..483H} to provide some general guidance for our inputs.

\subsection{\label{sec:cre}CRE density}

The CRE density is first scaled according to the compressed thermal electron density, $n_e$, i.e., the CRE density follows the mass density predicted by a Sedov profile. As discussed in Sec.~\ref{sec:regular}, this is a reasonable approximation since we only need a strong compression at the shock, which the Sedov profile provides. We then include a sharp cutoff to set the CRE density to zero in the interior 80\% of the shell and thus the CREs are confined to the outer 20\% of the shell, which at the scale we are modelling is equivalent to 45 pixels or 1.8~pc. Besides being compressed, we consider how the CREs may also be distributed around the shell. We use the standard isotropic, quasi-parallel, and quasi-perpendicular injection recipes to model these distributions \citep[][and references therein]{Jokipii:1982jy,1989ApJ...338..963L,Fulbright:1990gu}. In the quasi-perpendicular case, the CRE density is scaled by $\sin^{2}\phi_{Bn2}$, where $\phi_{Bn2}$ is the angle between the shock normal and the post-shock magnetic field, and in the quasi-parallel scenario the CRE density is scaled by $\cos^{2}\phi_{Bn2}$. 

CREs are accelerated at the shock. In the cases where we include a turbulent component of the magnetic field, the CREs will be affected by the component of the field that is locally parallel or perpendicular to the shock normal, depending on the injection recipe. In our modelling this extends some distance inside the shell (20\% of the SNR radius). For the CREs that remain interior to the shock, our implicit assumption is that these CREs have not had time to (azimuthally) diffuse far away from the location where they were accelerated on the timescales for which we are observing the SNR. Using the scale of 0.04~pc/pixel and a shock velocity of 5000~km/s, it would take nearly 8~years for the shock to cross a single pixel of this model SNR. 

\section{\label{sec:result}Results and Discussion}

In Fig~\ref{fig:bfield}, we show the three linear combinations of $|\mathbf{B}_{\textrm{reg}}|$, $|\mathbf{B}_{\textrm{rdm}}|$, and $|\mathbf{B}_{\textrm{rad}}|$ that we model: one dominated by the random component, one dominated by the radial component, and one where the components are comparable. The results of the modelling are shown in Figs.~\ref{fig:i}-\ref{fig:frac-plot}.

An interesting outcome is that where quasi-parallel or quasi-perpendicular acceleration cases are used, the simulated observations reveal that an apparently ordered polarization morphology is created even if the input magnetic field is completely turbulent. In the quasi-parallel case,  the magnetic field is apparently radial. Similarly, in the quasi-perpendicular case, we observe what appears to be a tangential magnetic field.  This can be explained as a selection effect due to the distribution of the CREs. 

For example, if we are using the quasi-parallel injection recipe and the case of a completely random magnetic field, there will be points all over the sphere that have a parallel component and CREs would be distributed uniformly over the sphere. At the particular locations where the magnetic field is parallel to the shock normal, the CRE density will be amplified.  Where the magnetic field is perpendicular, the CRE density will be suppressed. So although on average the CREs are distributed uniformly, the value will be high only in the particular cells where the magnetic field is radial. Since synchrotron emission depends only on the magnetic field component that is in the plane of the sky, this will be preferentially bright at the points where the field is radial creating a radial appearance to the polarization vectors.  

Observationally, the radial appearance created by an intrinsically radial magnetic field is very similar to an intrinsically turbulent field with a quasi-parallel CRE acceleration mechanism. These models are very similar to the observations of the young SNRs shown in Fig.~\ref{fig:historical_snrs}.

The polarized fraction is one possible observable that may help distinguish between these scenarios. It is very high for an intrinsically radial field but is greatly reduced when a random turbulent component dominates due to depolarization effects. In Fig.~\ref{fig:frac-plot}, we plot the mean and maximum polarized fraction in an annulus (with an inner radius set to 90\% of the SNR radius) for each of 20 azimuthal bins. We measure $\phi$ counterclockwise where $\phi=0$ is at the standard 3 o'clock position. We compare our models to SN1006 data from \citet{Reynoso:2013tr}, which has been rotated such that the bright limbs correspond to $\phi=90^\circ$ and $\phi=270^\circ$. We show the data for reference, but we do not attempt to fit a best model to the data. In practical terms, this is a subtle effect, especially considering that the foreground interstellar medium is also a source of depolarization.

An additional possible observable is polarization from dust. If the polarized dust emission is from thermal emission and not synchrotron, then the dust grains should be aligned with the intrinsic magnetic field and would not be impacted by the CRE acceleration mechanism. To date, there has only been one observation of dust polarization in a young SNR with a radial polarization morphology in the radio, which was for the case of Cas~A \citep{2009MNRAS.394.1307D}. These observations covered only a portion of the remnant and were of relatively low resolution (18'') compared to the size of the SNR (5'). These data nonetheless reveal that the dust polarization follows a similar radial pattern as the polarized radio emission, implying the existence of at least some fraction of an intrinsically radial component. 

An interesting case is that of a radially dominant magnetic field coupled with a quasi-perpendicular CRE acceleration mechanism (see Figs.~\ref{fig:i}-\ref{fig:frac-plot}, bottom-centre). In this case, we see evidence of the ordered compressed field, even when its amplitude is very small compared to the other components, not unlike the tangential field observed at the edges of SN1006 and Cas~A. Also like SN1006, this particular model reveals a bilateral morphology, magnetic field lines through the centre of the SNR being parallel to the axis of symmetry, and the polarized fraction is highest at the points where the total intensity is lowest \citep[see Fig.~4 in][]{Reynoso:2013tr}. In addition, when the strength of the turbulent component is comparable to the intrinsically radial component, as might be expected in the youngest SNRs, the bilateral morphology disappears (see Figs.~\ref{fig:i}-\ref{fig:frac-plot}, bottom-right). 

\section{\label{sec:conclusions}Conclusions}
We consider the origin of radial magnetic fields in young SNRs with the inclusion of the most commonly considered CRE acceleration mechanisms: quasi-parallel, quasi-perpendicular, and isotropic. We make the following conclusions:

\begin{enumerate}

\item We demonstrate that an intrinsic radial field is not required to produce an observed radial polarization, but that a turbulent magnetic field can create a radial appearance due to a selection effect that is dependent on the distribution of the CREs (i.e., using the quasi-parallel case).

\item Two observables that can potentially distinguish between an intrinsic radial field and a purely turbulent field are the fractional polarization and dust polarization.

\end{enumerate}

The case that is dominated by an intrinsically radial component and uses a quasi-perpendicular acceleration mechanism (Figs.~\ref{fig:i}-\ref{fig:frac-plot}, bottom-centre) is particularly interesting leading to an alternative explanation and several conclusions specific to this case:

\begin{enumerate}
\item The ordered compressed field can be revealed at the perimeter of the SNR, even when its amplitude is small compared to the amplitude of the other components.

\item Previous studies \citep[e.g.,][]{2017A&A...597A.121W}, which did not consider a radial component, were not able to model a radial appearance using quasi-perpendicular CRE geometry, whereas this case does.

\item This is the only example of bilateral morphology in the cases studied here, and we could find no case where a quasi-parallel CRE acceleration mechanism combined with an intrinsically radial field can have a bilateral appearance.

\item The observed magnetic field vectors through the centre of the SNR appear parallel to the axis of symmetry, which is consistent with what is observed in SN1006.

\end{enumerate}

The leading interpretation of SN1006 is a polar cap geometry with the radial field interpreted as the magnetic field lines converging at these caps. The review by \cite{2017arXiv170202054K} concludes that the majority of studies support this picture and therefore support the quasi-parallel CRE acceleration scenario. We suggest that the range of possible scenarios that can lead to a radial-looking magnetic field in SN1006 have not been fully explored. The current picture describing this SNR may be incomplete, especially considering that all young SNRs exhibit radial fields and that the origin of the radial nature is likely similar.

We show here the unanticipated result that quasi-parallel acceleration can create an apparently ordered radial component in a turbulent field through a selection effect. This effect may reinforce an existing intrinsic radial component.  

Alternatively, the case of an intrinsically radial field with a quasi-perpendicular CRE acceleration mechanism reveals intriguing similarities to SN1006 and the fractional polarization data would appear to support this scenario for this SNR. Further observations, such as high-frequency observations of SN1006 that would have less depolarization and more detailed modelling are required to distinguish between these scenarios.

\acknowledgements
The Dunlap Institute is funded through an endowment established by the David Dunlap family and the University of Toronto. J.L.W. and B.M.G. acknowledge the support of the Natural Sciences and Engineering Research Council of Canada (NSERC) through grant RGPIN-2015-05948, and of the Canada Research Chairs program. S.S.H. acknowledges the support of the NSERC (Discovery Grants and Canada Research Chairs programs), CSA and CFI. We thank L. Rudnick and E. Reynoso for providing data, and the anonymous referee for helpful comments that improved the manuscript.
\\
This research has made use of the NASA Astrophysics Data System (ADS).

\bibliography{references}

\end{document}